\documentclass[traditabstract]{aa}  

\usepackage{graphicx}
\usepackage{txfonts}
\usepackage{natbib}
\begin{document}

   \title{The Software Package for Astronomical Reductions with KMOS: SPARK}


   \author{R. I. Davies\inst{1}
\and
A. Agudo Berbel\inst{1}
\and
E. Wiezorrek\inst{1}
\and
M. Cirasuolo\inst{2}
\and
N. M. F\"orster Schreiber\inst{1}
\and
Y. Jung\inst{4}
\and
B. Muschielok\inst{3}
\and
T. Ott\inst{1}
\and
S. Ramsay\inst{4}
\and
J. Schlichter\inst{3}
\and
R. Sharples\inst{5}
\and
M. Wegner\inst{3}
   }

   \institute{Max-Planck-Institut f\"ur extraterrestrische Physik, Postfach 1312, 85741, Garching, Germany
     \and
   Institute for Astronomy, University of Edinburgh, Royal Observatory, Edinburgh, EH9 3HJ, UK
   \and
   Universit\"ats-Sternwarte M\"unchen, Scheinerstr. 1, 81679, M\"unchen, Germany
   \and
   European Southern Observatory, Karl-Schwarzschildstr. 2, 85748, Garching, Germany
   \and
 Department of Physics, University of Durham, South Road, Durham, DH1 3LE, UK}

   \date{Received ...; accepted ...}

 
  \abstract
   {KMOS is a multi-object near-infrared integral field spectrometer with 24 deployable cryogenic pick-off arms. Inevitably, data processing is a complex task that requires careful calibration and quality control. In this paper we describe all the steps involved in producing science-quality data products from the raw observations. In particular, we focus on the following issues: 
(i) the calibration scheme which produces maps of the spatial and spectral locations of all illuminated pixels on the detectors;
(ii) our concept of minimising the number of interpolations, to the limiting case of a single reconstruction that simultaneously uses raw data from multiple exposures;
(iii) a comparison of the various interpolation methods implemented, and an assessment of the performance of true 3D interpolation schemes;
(iv) the way in which instrumental flexure is measured and compensated.
We finish by presenting some examples of data processed using the pipeline.}


   \maketitle
%

\section{Introduction}
\label{sec:intro}

\begin{figure*}
\begin{center}
\resizebox{16cm}{!}
{\includegraphics{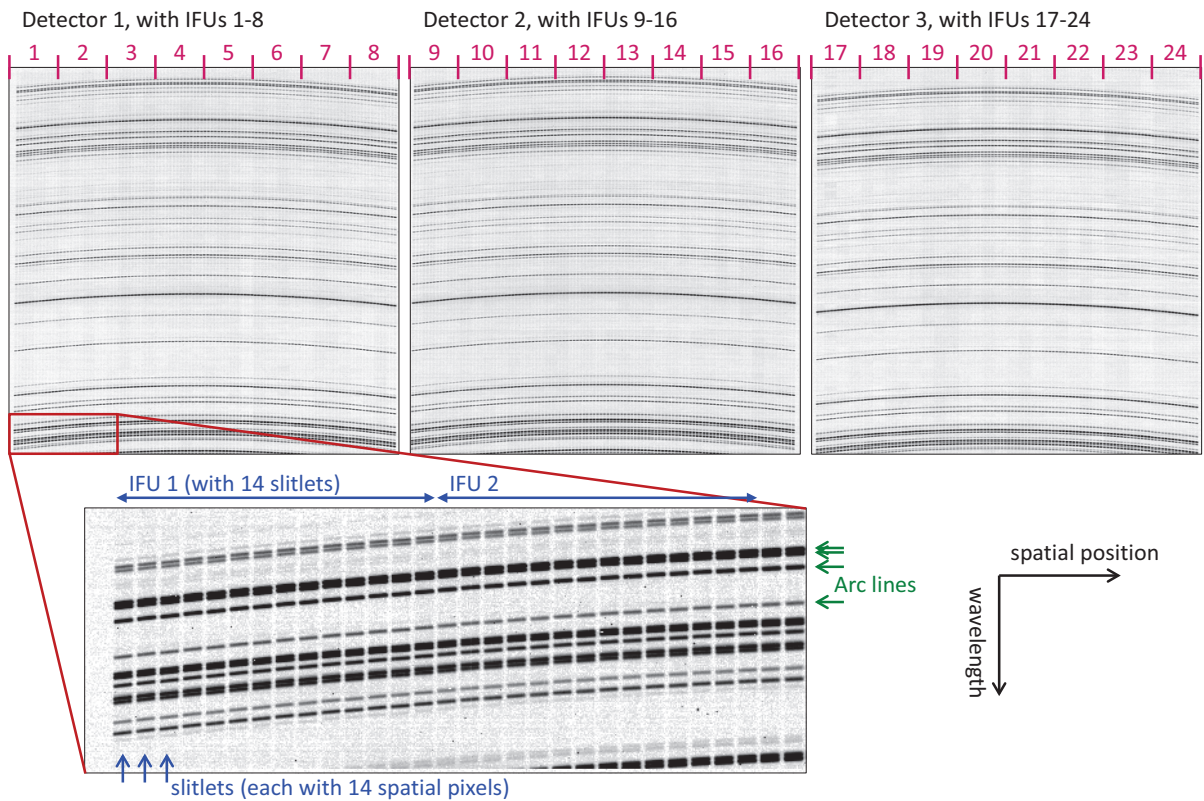}}
\end{center}
\caption{Images of H-band arc lines for the three segments, illustrating how the data are arranged on the detectors. The dispersion axis is approximately vertical, with longer wavelengths at the top and shorter wavelengths at the bottom. The horizontal axis denotes spatial position.
The curved black lines across each detector are the arc lines. Closer inspection shows that these are split into slitlets of 14 pixels, with 14 slitlets making up each IFU, and 8 IFUs across each detector.}
\label{fig:datalayout}
\end{figure*}

The K-band Multi Object Spectrograph (KMOS) is a fully cryogenic multi-object integral field spectrometer designed for seeing-limited operations \citep{sha06_newar}. 
Constructed by a consortium of German and British institutes, together with the European Southern Observatory (ESO), as one of the second generation Very Large Telescope (VLT) instruments, it has now been installed at a Nasmyth focus of the Unit Telescope 1 and commissioned \citep{sha13_mess}.
It is equipped with 24 integral field units, each with a $2.8\arcsec\times2.8\arcsec$ field of view sampled at 0.2\arcsec.
These are deployed on-the-fly by robotic arms, to positions within the 7.2\arcmin\ patrol field that have been previously allocated by the observer in a configuration file.
The number and size of the integral field units (IFUs), and the spectral resolution, have been chosen according to the key science driver \citep{sha05_mess}, to study galaxy evolution through cosmic time, which can be achieved by mapping the kinematics and morphology of large samples of high redshift galaxies.
The instrument has been designed with three quasi-identical segments, each of which comprises 8 pick-off arms, a filter wheel, an integral field unit (which uses mirrors to slice the images and re-arrange the pieces along a pseudo-slit), a spectrograph, and a $2K\times2K$ HAWAII-2RG detector.
Each segment has five gratings that together span the 0.8--2.5\,$\mu$m range.
Four of these (IZ, YJ, H, and K) each cover a single broad band at a resolving power of $R\sim4000$, and the fifth (HK) covers the H and K bands simultaneously at a lower resolution.
As indicated by Fig.~\ref{fig:datalayout}, segment~1 is associated with IFUs 1--8, segment~2 with IFUs 9--16, and segment~3 with IFUs 17--24.

Despite the opto-mechanical complexity of KMOS, it is a relatively simple instrument to configure and use because there is only one observing mode.
The only complex task, allocating the 24 arms to specific targets in the patrol field, is handled by the KMOS Arm Allocator (KARMA; \citealt{weg08_spie}), an automated tool that optimises the assignments taking into account the mechanical constraints of each arm (i.e. the limits to which specific arms can reach) as well as target priorities.
One can allocate IFUs to science targets, reference objects (e.g. stars to monitor variations in throughput, seeing, etc.), or sky.
By having small dithers between exposures, and also switching between 2 pointings, one can alternate IFUs between object and sky positions at a user-specified frequency.
Alternatively, it is possible to activate the mapping mode, which uses either 16 pointings to map out a contiguous $65\arcsec\times43\arcsec$ field using all 24 arms, or 9 pointings to map out a $32\arcsec\times16\arcsec$ field using 8 arms.

\begin{figure*}
\begin{center}
\resizebox{16cm}{!}
{\includegraphics{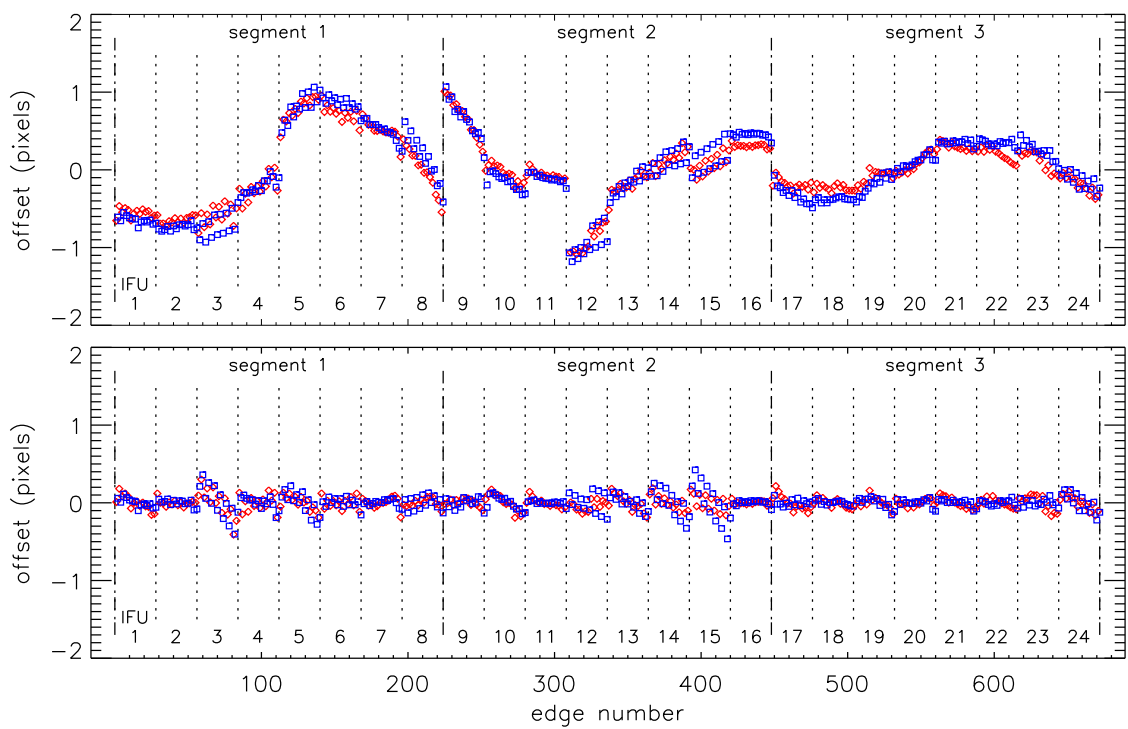}}
\end{center}
\caption{Offsets between the measured location of each slitlet edge and its ideal position (for a specific band, rotator angle, and instrument temperature). The locations are calculated from the polynomial fits to each edge, for a single row of pixels across the detectors. Left edges are plotted as red diamonds, right edges as blue squares. The dotted lines indicate the separation between IFUs, and the dashed lines between the 3 instrument segments. An offset of 1 pixel is equivalent to 18\,$\mu$m at the detector.
Top panel: offsets calculated for the ideal case that the slitlet widths and spacings should be perfectly uniform across each segment (i.e. how far the edges differ from the location expected, based on the average slitlet width and spacing on that detector).
Bottom panel: offsets derived for each IFU individually, i.e. based on the mean slitlet width and spacing per IFU. The distinct pattern (e.g. in IFU 3) is a direct imprint of the opto-mechanics.}
\label{fig:edge_pos}
\end{figure*}

\begin{figure*}
\begin{center}
\resizebox{16cm}{!}
{\includegraphics{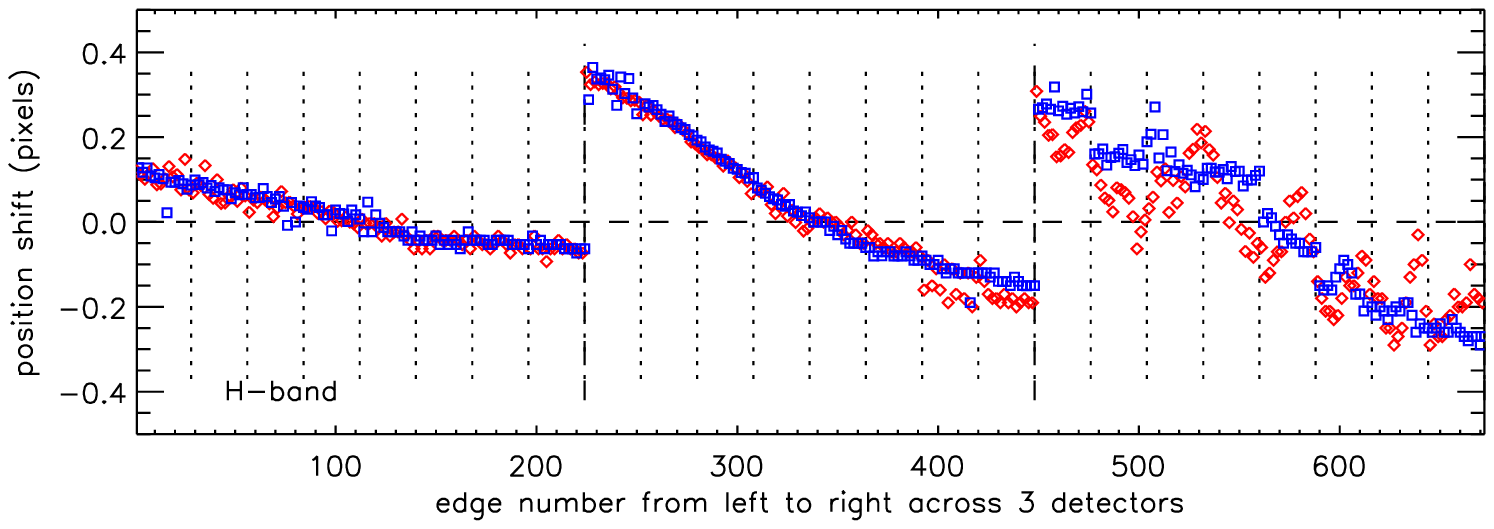}}
\end{center}
\caption{Shifts of slitlet edge positions (for the same band and rotator angle) when the temperature measured in the cryostat has changed by 2\,K.
Constant offsets have been removed, so that only the impact of stretching is shown.
Left edges are plotted as red diamonds, right edges as blue squares. The dotted lines indicate the separation between IFUs, and the dashed lines between the 3 instrument segments.
The pattern of data, as projected onto each detector, spans a width of about 2007\,pixels.
This figure shows that there is a differential shift of $\sim$0.2--0.4\,pixels in the location of the left-most and right-most edges on each detector; corresponding to a stretch of $\sim$0.01--0.02\% in the projected size of the pattern.
While apparently a very small effect, this has a serious impact on the data calibration.
The global shape in the figure probably originates in the spectrographs, while the more complex (intra-IFU structure) is likely associated with the image slicing optics.}
\label{fig:temp_shift}
\end{figure*}

In terms of the layout of the data, the $14\times14$ spatial pixels\footnote{Depending on context, the term `pixel' can be used to denote either the distance between sampled points on a detector, or the distance between interpolated grid points in a reconstructed cube. These are not necessarily the same, since the pipeline allows the spatial grid spacing in the cubes to be freely specified.} from each of the 24 IFUs are arranged together as a series of 336 dispersed slitlets across the 3 detectors, with the spectral axis approximately aligned with pixel columns on the detectors.
This is illustrated in Fig.~\ref{fig:datalayout}.
We begin this paper by looking in Section~\ref{sec:edges} at how the positions of the edges of these 336 slitlets can be used as a diagnostic of the instrument alignment, flexure, and repeatability.
The edge positions are also a key part of the instrument calibration.
This is discussed in Section~\ref{sec:calib}, where we describe our concept for the calibration frames and how we have implemented it.
Section~\ref{sec:interp} turns to the reconstruction of the data cubes.
We describe how we have minimised the number of interpolations, and compare the various interpolation methods available in the pipeline.
The impact of spatial and spectral flexure, and the ways in which they can be measured and compensated, are the topic of Section~\ref{sec:flex}.
Section~\ref{sec:workflow} outlines the main steps in processing science observations.
Finally, in Section~\ref{sec:examples} we present some examples of data that have been processed with the pipeline, and conclude with Section~\ref{sec:conc}.

\section{Slitlet Edge Locations}
\label{sec:edges}

The 672 slitlet edges, with a measurement repeatability of 0.02\,pixels, provide a remarkably detailed diagnostic of the instrument's optical alignment via their imprint on how the data is projected onto the detectors. This is important for the calibrations (which are described in Section~\ref{sec:calib}), and also yields insights about the instrument itself.
These not only help to compensate flexure (see Section~\ref{sec:flex}), but reflect the optical design and manufacturing quality of the opto-mechanics.

An example is given in Fig.~\ref{fig:edge_pos}, which shows how much the actual measured location of each edge differs from its ideal position (i.e. where it would be if the slitlet widths and spacings were exactly uniform within any given segment).
To do this, the locations of the edges were calculated from their polymonial fits across a single row of pixels in the middle of each detector.
Based on the slope of the slitlet traces, calculating edge locations along a single row rather than following a specific wavelength, will lead to an offset of not more than 0.25\,pixels, with a symmetric shape across each segment.
Instead, the global shape is approximately anti-symmetric, with an amplitude of 1\,pixel.
It likely arises within the spectrograph, while the details originate from the image slicer.
The mean slitwidth is 13.6\,pixels and has a standard deviation between IFUs of 0.1\,pixels.
This small variation in slitlet width is compensated by their separation, so that the mean distance between left or right edges within any IFU is always almost exactly 18\,pixels.
The figure also demonstrates the precision of the image slicing opto-mechanics: in most cases, the slitlets within an IFU are positioned on the detector to a relative precision of $\pm$1\,$\mu$m.
In a few cases, one can see the imprint of the opto-mechanics design (most obviously in IFUs 3 and 15). 
This pattern arises from the pupil mirror array between the slicing mirrors and the slit mirrors, which is manufactured as two rows of 7 mirrors slightly offset from each other \citep{sha06_newar,sha06_spie}.

A second example of how the slitlet edges can be used to track changes in the instrument is given in Fig.~\ref{fig:temp_shift}.
This is related to the temperature of the opto-mechanics inside the cryostat.
Although the instrument is cooled, only the temperature of the detectors is controlled; the temperature of the optical bench can vary.
The figure shows how much the relative position of each edge shifts when the cryostat temperature increases from 118\,K to 120\,K, and thus that temperature variations can impact the pattern of the data as projected onto the detectors.
The differential shifts between the left-most and right-most edges on each detector indicate that the width of the projected pattern of data on each detector stretches by about 0.1\,pixels per 1\,K temperature change.
Although this is opto-mechanically very small (0.1\,pixels corresponds to a relative stretch of only 0.005\%), it is important because a valid reconstruction requires the calibrated positions of the slitlet edges to match the positions during science observations.
The maximum tolerance on any edge is $\sim$0.2\,pixels, although a better match will yield better results.
As discussed in Section~\ref{sec:flex}, the operational implication is that the temperature of the cryostat during the calibrations must be within $\sim\pm$1\,K of that during the science observations.


\section{Calibration}
\label{sec:calib}

\begin{figure*}
\begin{center}
\resizebox{\hsize}{!}
{\includegraphics{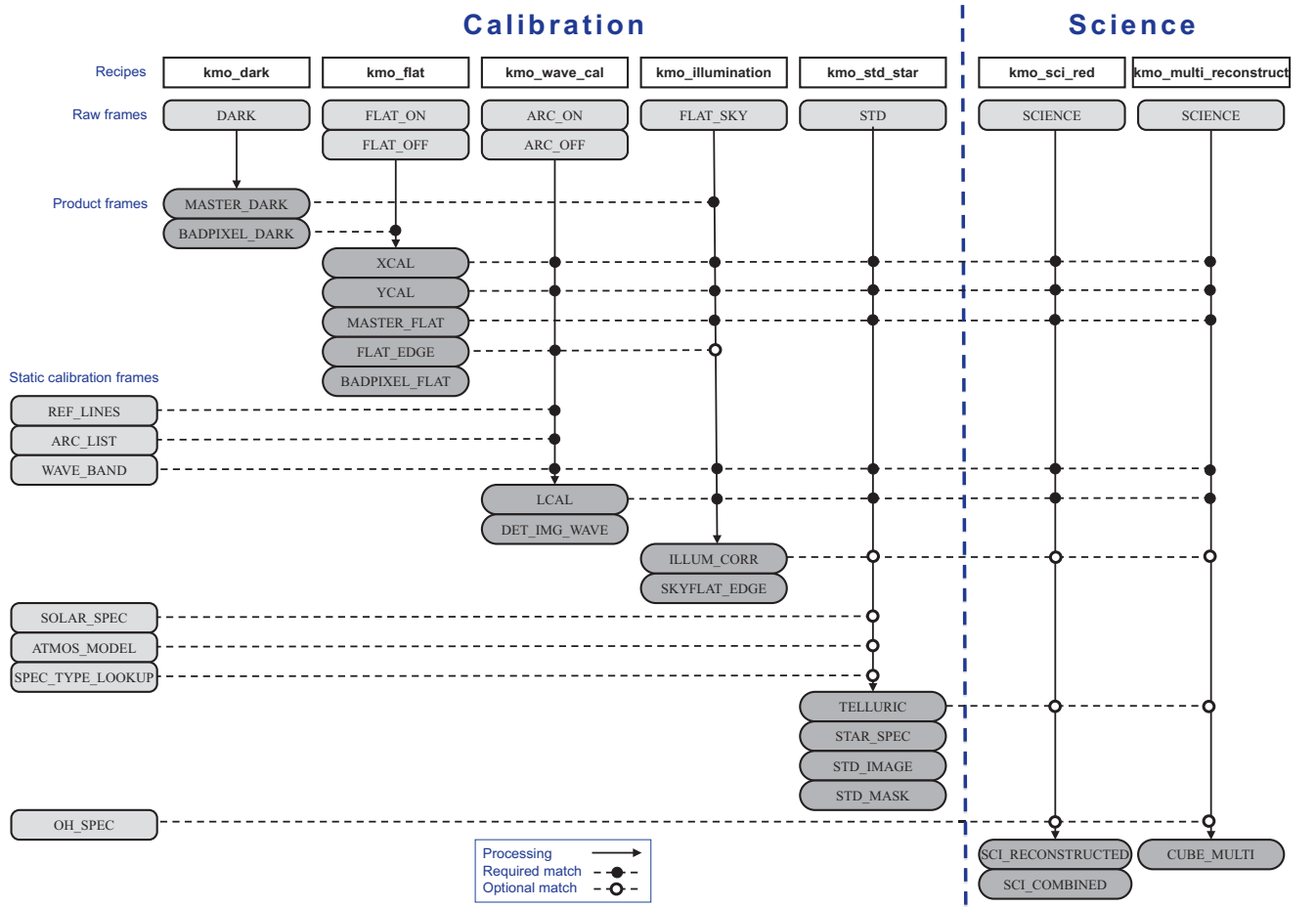}}
\end{center}
\caption{Calibration Cascade for KMOS. The top rows show the different recipes (modules) executed, and the raw files they require. Underneath are the products from each recipe. The horizontal lines showing how, together with some extra static calibrations, these feed into subsequent recipes, leading to the cascade.
Here, the instrumental calibrations (kmo\_dark, kmo\_flat, kmo\_wave\_cal, kmo\_illumination) and the astronomical calibration (kmo\_std\_star) have been separated from the recipes that process the science data itself (kmo\_sci\_red and kmo\_multi\_reconstruct).
}
\label{fig:cascade}
\end{figure*}

Instrumental and astronomical calibrations for KMOS can be assigned to three different purposes:
(i) flatfielding, i.e. correcting for differences in individual pixel gains on the detector and for spatial variations in throughput of the instrument;
(ii) mapping the sliced data into a regularly gridded cube; 
(iii) correcting for spectral variations in atmospheric and instrument transmission, and applying a flux calibration.
In this simple scheme, both (i) and (iii) are performed in a standard way for near-infrared data. In contrast our concept for the calibrations required for the interpolation in (ii) is a little different to many instruments, and is discussed in more detail below.

The three steps can be identified on the calibration cascade shown in Fig~\ref{fig:cascade}, which includes some additional details.
The pipeline modules (recipes) are listed across the top row, together with the raw data files they require.
Underneath, the products from each recipe are listed.
From left to right, these and some additional static calibrations, feed into the subsequent recipes in a cascade.

\subsection{Flatfielding}

Since KMOS has no imaging mode, variations in pixel gain are measured using the dispersed light from internal halogen lamps. 
Although the spectral shape of the lamps is removed using a running median along the dispersion axis, spatial variations in the flux (due to possible mis-registration of the arm reference positions for internal calibrations) are typically compensated separately using sky flats taken during twilight.
When these latter exposures are processed, the internal flatfield is applied and then the data are reconstructed.
The resulting cubes are collapsed along the spectral axis to produce images that, by design, trace the spatial non-uniformity of the internal flatfields.
If the correction is not too large (i.e. the calibration reference positions for the arms are well aligned with respect to the internal lamps) it is instead possible to derive this illumination correction directly from the internal flatfields by spatially smoothing and inverting the reconstructed image of the flatfield.

\begin{figure*}
\begin{center}
\resizebox{12.5cm}{!}
{\includegraphics{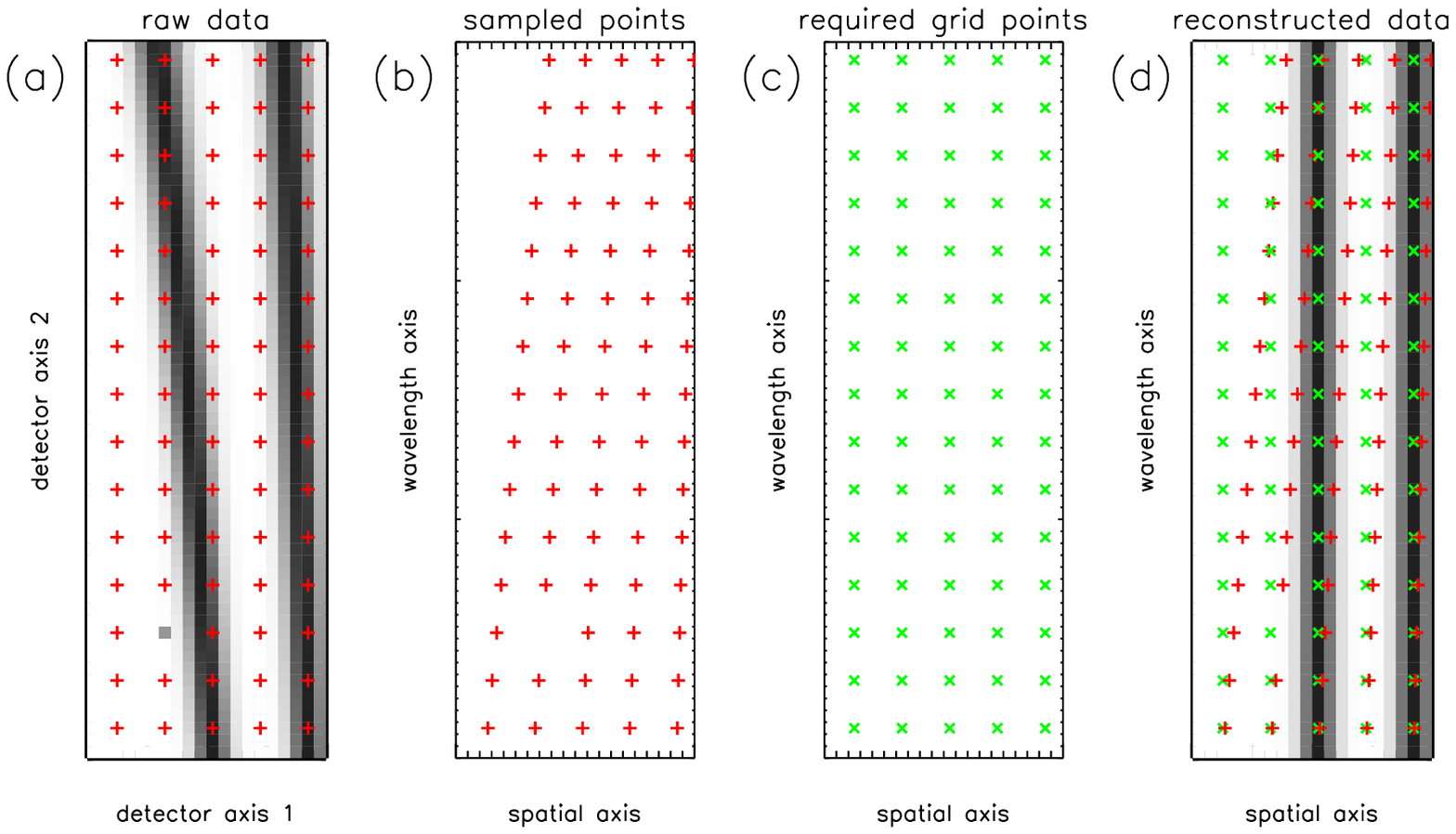}}
\end{center}
\caption{ \label{fig:interp} 
Interpolation scheme illustrated in 2-dimensions.
(a) observed data are sampled regularly in the reference frame of the detector (red points).
(b) this sampling is irregular in the reference frame of the reconstructed cube; bad pixels can simply be omitted from the set of valid samples. The calibration products directly provide the mapping from (a) to (b).
(c) one can freely specify the required gridding (i.e. spatial/spectral pixel scale) for the reconstructed data (green points); it is independent of the actual sampling on the detector, although the resolution one achieves is not.
(d) each required grid point (green) is interpolated from sampled points (red) which lie in its local neighbourhood. Any suitable algorithm can be used for the interpolation.
}
\end{figure*} 

The flatfield frames are also used to identify bad pixels.
This is a 2-step process.
First, a series of dark frames is used to identify hot pixels.
For dark exposures (not affected by persistence) of at least 60\,s, the number of hot pixels stabilises at $48/18/17\times10^3$ for the 3 detectors, corresponding to 0.5--1\% of the pixels.
The flatfields are then used to flag cold pixels.
These are identified by their low flux, and so include also pixels between the slitlets.
As such, the definition of `bad' pixels for KMOS is rather broader than usual, and should more realistically be considered simply as those not used during the reconstruction of the datacubes.
The final bad pixel mask is a combination of the pixels identified from the dark and flatfield frames respectively.
Values for these pixels are not interpolated from their neighbours; instead they are simply omitted from the look-up tables derived from the calibration frames, and hence ignored during the reconstruction, as described below.

\subsection{Mapping Distortions}
\label{sec:mapdistort}

Interpolation is a crucial issue for integral field spectroscopy, and poor management of the interpolation strategy can degrade the final data quality.
For this reason, the perspective adopted for the KMOS pipeline enables the interpolation to be performed in a single step, while at the same time still permitting flexibility.
In essence, the calibrations allow one to create look-up tables that directly associate each measured value on the detector with its spectral and spatial location in the final reconstructed data. 
The strengths of this strategy can be summarised as follows:
\begin{enumerate}[(i)]
\item
bad pixels are simply ignored because they do not appear in the look-up
tables, 
\item
one has the freedom to choose any spectral and spatial sampling
in the processed product,
\item
rotation of the instrument during the observations can be accounted for during the reconstruction,
\item
spatial and spectral flexure can be corrected by applying appropriate adjustments to the look-up tables before reconstructing the data, and
\item
by including multiple exposures in the look-up table, one can
combine and interpolate all the data on a given object simultaneously in one step (taking into acount different flexures and rotation angles),
\item
rather than reconstructing the data, it is possible instead to map a 3D model of an object observed back to the detector plane, i.e. onto data that have not been resampled.
\end{enumerate}

The scheme is outlined graphically in Fig.~\ref{fig:interp}.
The most important realisation is that in `detector space' there can
be no concept of a wavelength or spatial axis.
These concepts apply only to processed data.
The detector is nothing more than the medium on which raw data values are
recorded.
The calibrations allow one to assign each measured value on the
detector to a spatial/spectral location in the
reconstructed cube.
Together, these locations provide an irregularly spaced sampling of
that cube.
The aim is thus to reduce the raw data and the calibrations to
a list of sample values with their associated locations in the cube:
\[
\begin{array}{cccc}
value_0, & x_0, & y_0, & \lambda_0 \\
value_1, & x_1, & y_1, & \lambda_1 \\
\vdots & \vdots & \vdots & \vdots \\
value_n, & x_n, & y_n, & \lambda_n \\
\end{array}
\]

In practice, this is done by creating a set of three frames which together provide the $(x,y,\lambda)$ location in the final cube for each valid pixel on the detector (where $x$ and $y$ are the horizontal and vertical spatial directions within an IFU, corresponding to east and north if the rotator offset angle is zero).
These frames are, literally, look-up tables stored in image format.
One additional requirement for KMOS, which has 24 IFUs, is to identify to which IFU each value belongs.
Because each spatial location is given as an integer in milliarcsec from the centre of an IFU, the IFU identification itself is encoded as a number after the decimal point in those frames.
Any interpolation scheme -- not just those implemented in the pipeline, which are discussed in Section~\ref{sec:interp} -- can easily make use of these calibration frames.
However, it is important to bear in mind that the calibration frames created and saved to files by the pipeline are generated without any flexure correction. 
The corrections are applied only internally within recipes and are tailored each time to match the specific science frames being processed.

\begin{figure}
\begin{center}
\resizebox{\hsize}{!}
{\includegraphics{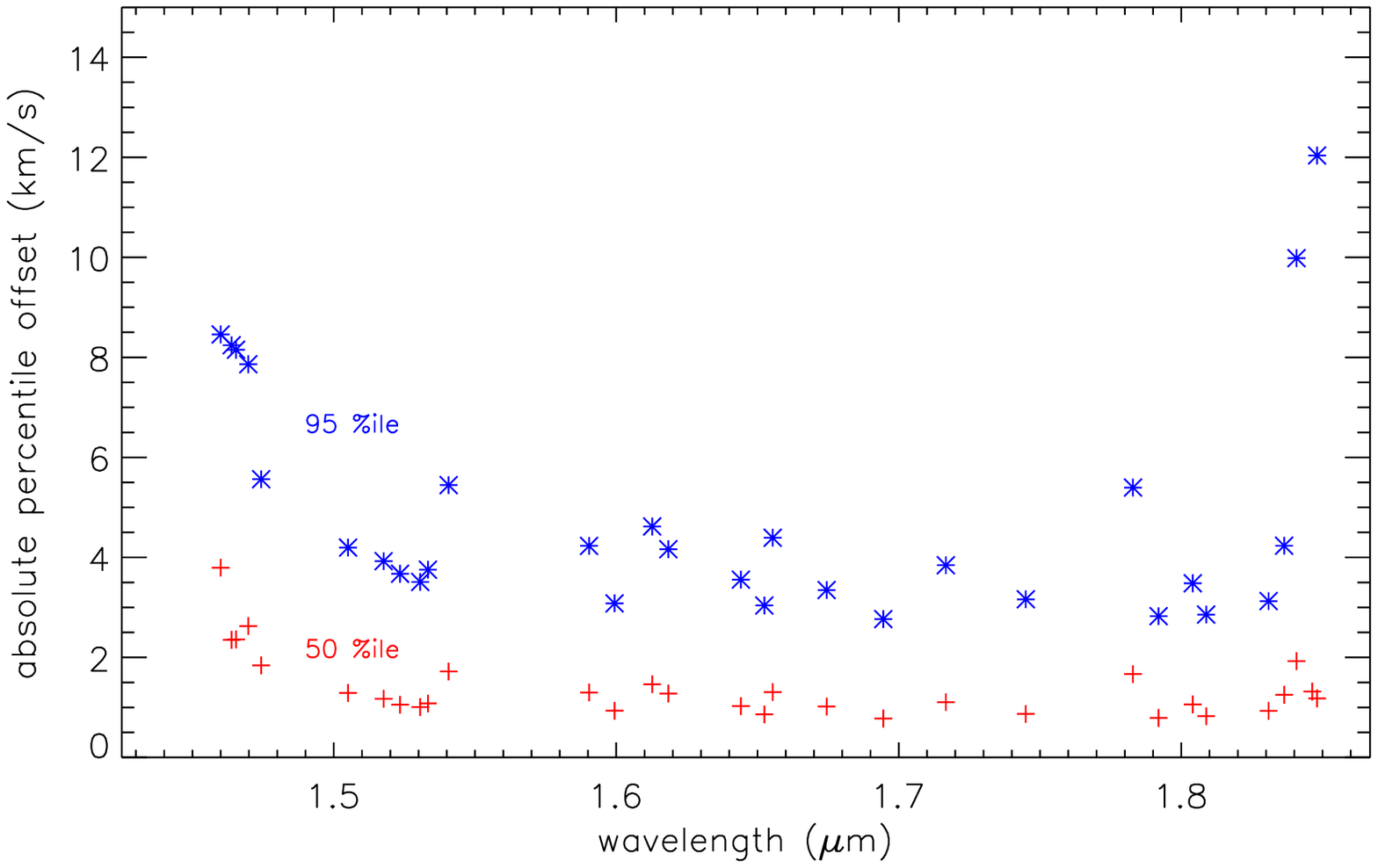}}
\resizebox{\hsize}{!}
{\includegraphics{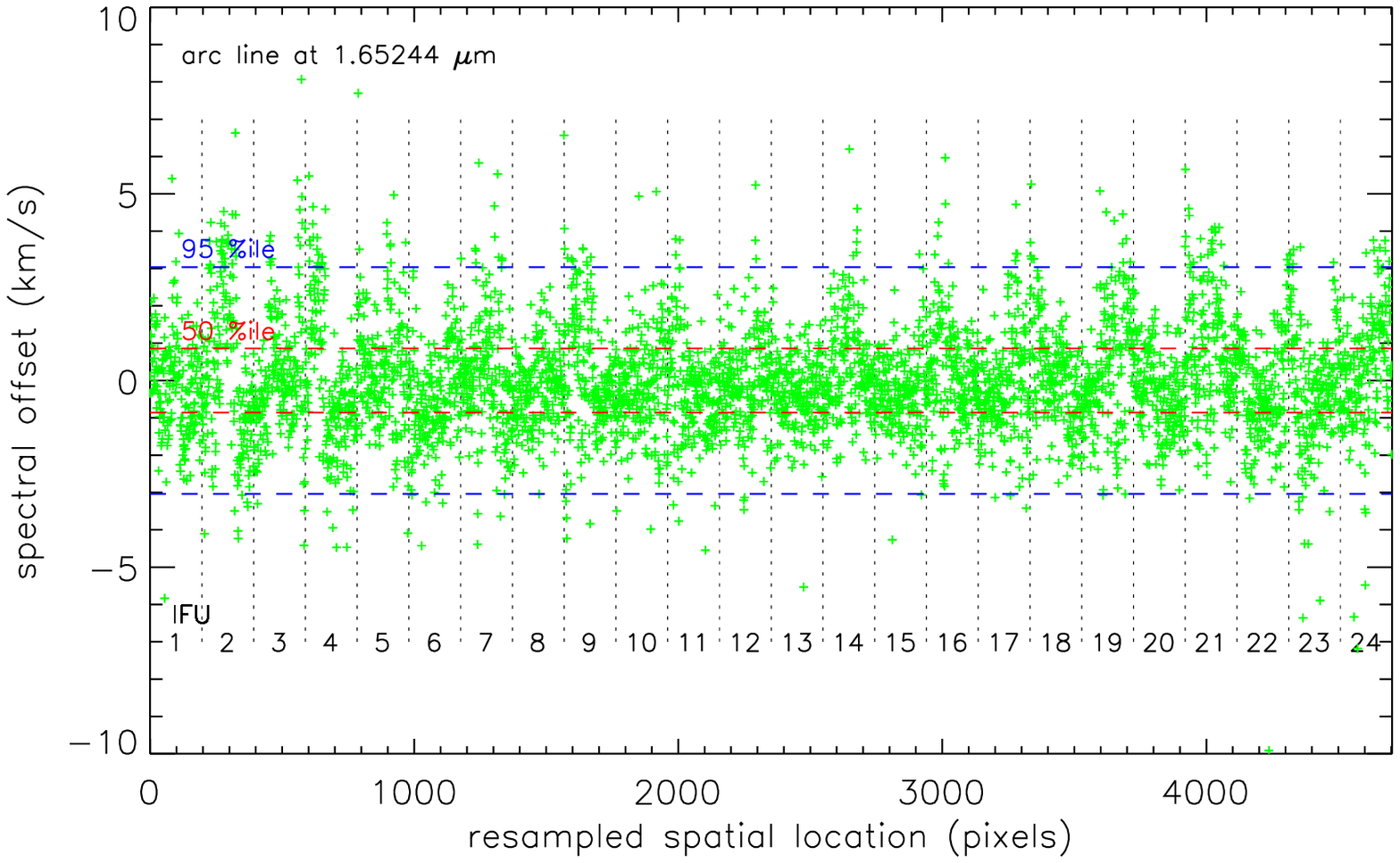}}
\end{center}
\caption{\label{fig:hqual}
Measurement of the quality of the wavelength calibration from a reconstructed H-band arc-line frame.
Top: for each arc line, the wavelengths are measured at all spatial locations in the 24 reconstructed cubes. The red and blue points then denote the 50-percentile and 95-percentile absolute deviations of the measurements from the expected wavelengths.
Bottom: for a single line at a central wavelength, the points (green) show the deviation for each spatial location. The dashed red and blue lines show the 50-percentile and 95-percentile respectively.}
\end{figure}

There are two key steps when creating the calibration frames.
The first is detecting and measuring both the rising and falling edges of the slitlets in the flatfield frames.
The basic process is to fit a Gaussian to the gradient of each edge, to identify its location, and then fit a polynomial function to the resulting set of points along the dispersed length of the slitlet.
These edge functions form the basis for specifying the location of each slitlet on the detectors, and for assigning pixels a spatial position within the IFU field of view -- which is done under the assumption that, projected on the sky, the slitlets are $0.2\arcsec\times2.8\arcsec$, and that pixels are distributed uniformly (linearly) between the slitlet edges.
The wavelength solution is found by first identifying a few relatively isolated lines in the arc-lamp spectrum, in order to have a rough estimate of where to look for other lines.
The lines ($\sim30$ or more per band) are traced across each slitlet, and then the wavelengths at each pixel are assigned by fitting a polynomial along the dispersion direction.
Typically, a 6th order function is used since this fits the data well at all spatial locations (although lower order functions may still yield comparable wavelength solutions across most spatial locations).
An example of the quality of the calibration is shown in Fig.~\ref{fig:hqual} for the H-band.
This shows the 50-percentile and 95-percentile absolute offsets of each arc line as measured over all spatial locations of the 24 cubes in a reconstructed frame.
These are $\sim$1\,km\,s$^{-1}$ and 3--4\,km\,s$^{-1}$ respectively, confirming that, with good signal-to-noise, the wavelength calibration itself (excluding any flexure issues, see Section~\ref{sec:flex}) is precise to a few percent of the line width, and well within the specification of 10\,km\,s$^{-1}$.

When performing the calibration, each slitlet is treated independently, to limit the order of the fit along the spatial axis, and because there can be discrete shifts between them due to the finite optical manufacturing tolerances.

\subsection{Standard Stars}

\begin{table*}
\caption{Parameters for KMOS flux calibration. Data for the 2MASS bands are from \cite{coh03}.}
\label{tab:phot}
\begin{tabular}{cclll}
\hline\hline
KMOS & 2MASS & band pass       & \multicolumn{2}{c}{zero magnitude flux density}    \\
band & band & $\mu m$ & $W m^{-2} \mu m^{-1}$ & $ph s^{-1} m^{-2} \mu m^{-1}$\\
\hline
K &  K & 2.038--2.290 & $4.28\times10^{-10}$ & $4.65\times10^9$\\
HK & H \& K & 1.5365--1.7875 \& 2.208--2.290 & $1.13\times10^{-9}$ \& $4.28\times10^{-10}$ & $9.47\times10^9$ \& $4.65\times10^9$ \\
H & H & 1.5365--1.7875 & $1.13\times10^{-9}$ & $9.47\times10^9$ \\
YJ & J & 1.154--1.316 & $3.13\times10^{-10}$ & $1.94\times10^9$ \\
IZ & --- & 0.985--1.000 & $7.63\times10^{-9}$ & $3.81\times10^{10}$ \\
\hline
\end{tabular}
\end{table*}

As is usual for near-infrared spectroscopy, the same standard stars are used both for telluric correction and flux calibration.
The operating procedure adopted for KMOS allows one to use any of the very large number of bright G2V or OBA stars with known spectral type and magnitude.
The data are fully processed up to the point at which a reconstructed cube is available.
From this, the pipeline will extract an integrated spectrum and attempt to correct it for stellar features (either using a solar spectrum, or by fitting out the most common lines) in order to generate a telluric spectrum.
An atmospheric model is used to guide the line fitting, since some stellar lines are blended with atmospheric lines.
However, the implementation of this stage is very simple and so its success can vary greatly depending on the input data, the stellar type, the waveband, whether the stellar lines themselves are blended together (OBA stars), and the wavelength coverage of the solar spectrum (G2V stars).
So there is always the option for the user to do these last steps manually, using more sophisticated tools that have been developed specifically for this purpose.
In paticular, for early A-type stars, \cite{vac03} developed a very successful tool that is able to derive the telluric correction for medium resolution spectra across the 0.8-2.5\,$\mu$m range using a high resolution model of an A0~V star.
Given the availability of libraries of model spectra, at high resolution over a broad wavelength range, that cover a wide parameter range (e.g. \citealt{hus13}), one can consider applying a similar technique to stars of other spectral types.

\begin{figure}
\begin{center}
\resizebox{\hsize}{!}
{\includegraphics{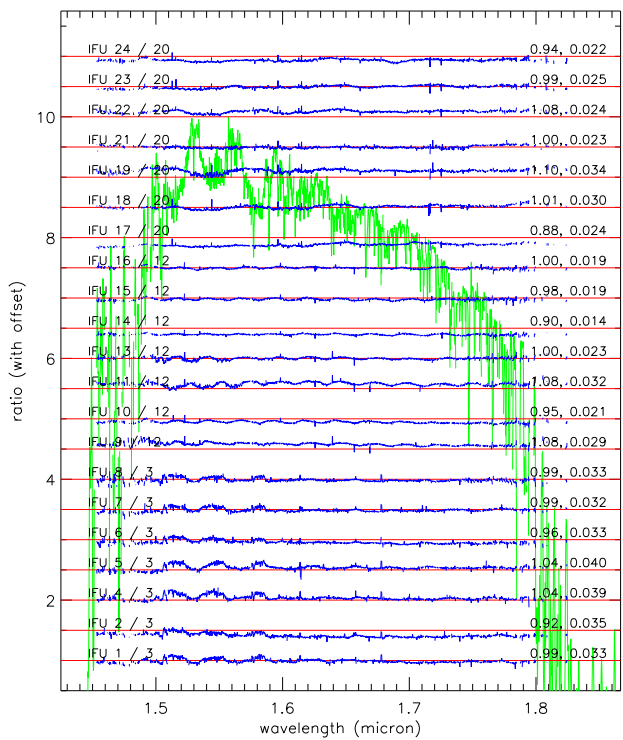}}
\end{center}
\caption{ \label{fig:stararms} 
Comparison of H-band spectra of the same star measured in just IFUs 3, 12, and 20, to spectra of the same star measured in the other IFUs.
The spectrum is shown in green in the background.
The blue lines show the ratio between the spectra in 2 different IFUs from the same instrument segment (with the red lines indicating a ratio of 1 for reference).
Note that one can see a ripple effect at 1.5--1.6\,$\mu$m, which is discussed in Section~\ref{sec:multirecon} and shown more clearly in Fig.~\ref{fig:wiggle} for the same dataset.
The numbers on the right are respectively, the mean and standard deviation of this ratio (for regions where the signal-to-noise is above a minimum threshold).
This suggests that using just 3 arms to calibrate all IFUs is sufficient at the level of precision of a few percent.
}
\end{figure} 

Flux calibration is more robust because of the availability of photometry from 2MASS.
As such, the 2MASS passbands \citep{coh03} are applied when calculating the zeropoints of KMOS for the YJ, H, HK, and K bands.
Similarly, the 2MASS zero magnitude flux densities are used for throughput estimates or for deriving line fluxes.
Since the $z$ band is poorly defined, for the KMOS IZ band a pseudo-monochromatic 1\,$\mu$m flux density is used.
The appropriate stellar magnitude can be interpolated from the KHJIR magnitudes, where the latter two come from the USNO-B catalogue \citep{mon03}.
The parameters used for KMOS are summarised in Table~\ref{tab:phot}, indicating the 2MASS bands for which magnitudes should be used to calibrate the various KMOS bands.
The zeropoint, which is written into the header of the processed files, is defined in the usual way so that
\[
magnitude = zeropoint - 2.5log_{10}(counts/second)
\]
and is calculated for each IFU separately.
The pipeline uses the total flux within the field of view, as is appropriate given the limited field of each IFU and the typical seeing.

One final, but important, consideration is whether one needs to measure a telluric correction spectrum for each arm?
The standard calibration procedure is to put the star in just one arm for each instrument segment, so that 3 telluric spectra are measured. 
Each one is then used for all arms in that same segment.
In most cases, this is likely to be sufficient; but if more precise telluric correction is required, it is possible to request that the star be observed in every arm, although this takes significantly longer.
Fig.~\ref{fig:stararms} attempts to quantify the impact of using the standard calibration procedure by looking at the ratio of the spectra from the same star observed with different arms in the same segment.
Across all bands, the typical standard deviation of the ratios is a few percent, suggesting the calibration can be done to that level of precision with the faster 3-arm standard scheme rather than needing to dither the star through all 24 arms, which can take 10--20\,mins longer.

%

\section{Interpolation and Reconstruction}
\label{sec:interp}

As discussed in Section~\ref{sec:mapdistort}, there are 3 calibration products that together contain the $(x,y,\lambda)$ coordinates for each illuminated detector pixel.
This is all the information required to reconstruct the datacubes in a single step, perhaps using just one 3D interpolation.
This section focusses on the algorithms available in the pipeline to do that.

\subsection{Interpolation Methods}

There are currently four different types of interpolation implemented in the pipeline.
A user is not necessarily limited to these methods, but can in principle also use the calibration frames with their own interpolation scheme.
It is important to realise that all the true 3D methods are weighted averages of neighbouring points, and so are limited in their performance by the sampling of the data.
The reason for this, and a way to overcome it, are described afterwards in Section~\ref{sec:multirecon}.

\subsubsection{Nearest Neighbour}

\begin{figure*}
\begin{center}
\resizebox{14cm}{!}
{\includegraphics{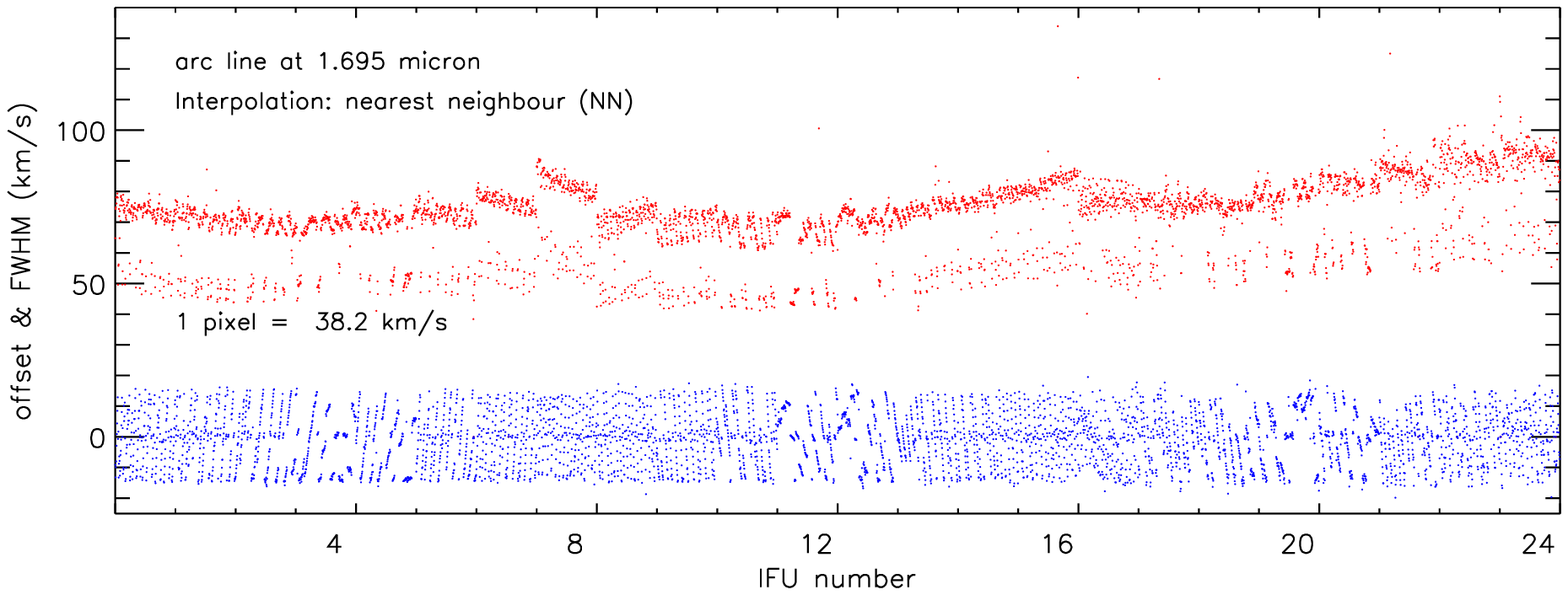}}
\resizebox{14cm}{!}
{\includegraphics{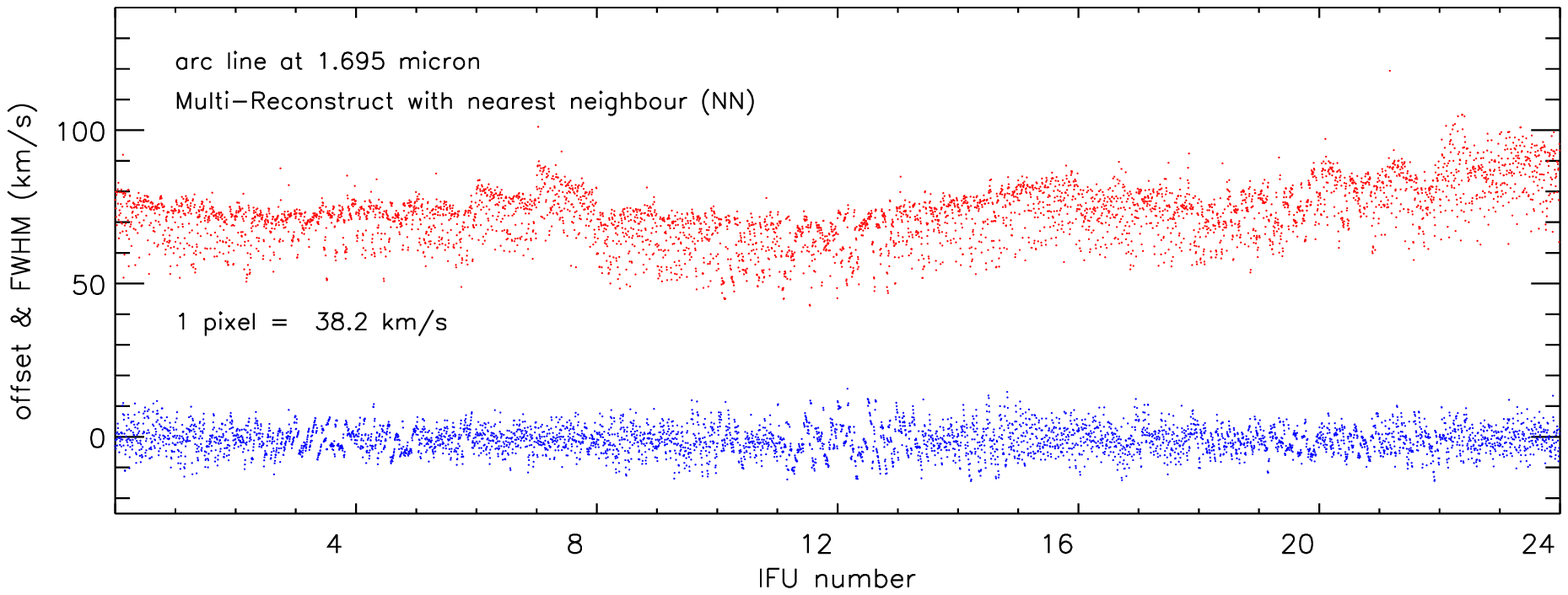}}
\end{center}
\caption{ \label{fig:nn} 
Spectral fidelity of an arc-line in the H-band as a function of spatial position across all 24 IFUs, when using the nearest neighbour reconstruction algorithm.
The offset of the line from its nominal position (blue), and its FWHM (red), are shown.
Top: for a cube reconstructed from a single exposure.
Bottom: using the multi-reconstruct scheme described in Section~\ref{sec:multirecon} to generate a single cube from 6 input frames at different rotator angles.
The improvement can be quantified in terms of the rms error in the line position, which is 8.6\,km\,s$^{-1}$ for the single frame, and 4.5\,km\,s$^{-1}$ when using 6 frames simultaneously.
}
\end{figure*} 

This is a fast, but rather approximate, method in which no interpolation is actually performed.
Instead, the data are simply re-arranged.
As such, it does not affect the noise properties.
However, it also does not provide the highest spatial or spectral resolution.
This is because any given value may be offset by up to 0.5\,pixels (i.e. sample points) in any of the 3 dimensions from where it originated, since this is how far away the neighbour can be.
The resulting poor fidelity\footnote{here, we use the term `fidelity' to denote the spectral quality in terms of the width and location of an emission line} is demonstrated for the spectral axis in the upper panel of Fig.~\ref{fig:nn}.
In this figure, the thickness of the blue band representing the wavelength offset reflects the finite sampling of the data: the points are distributed roughly uniformly within that range.
The global shape of the red band is due to the variation in instrumental resolution across each detector.
The apparent reversal of some IFUs is due to the ordering of the slitlets (due to the symmetry of the segments in KMOS, and the location of the pick-off arms in a circle around the patrol field, the ordering of the pixels and slitlets on-sky differs between groups of IFUs).
The brighter band at a FWHM of 70--90\,km\,s$^{-1}$ traces the native resolution of the sampled data.
The fainter red band at 50--60\,km\,s$^{-1}$ is due to the presence of some locations where the emission line is very nearly centered in the middle of a pixel spectrally.
In these instances, a Gaussian fitted to the line has a narrower width than the more common situation where the flux is spread more equally over 2 pixels.

\subsubsection{Linear Distance Weighting}

\begin{figure*}
\begin{center}
\resizebox{14cm}{!}
{\includegraphics{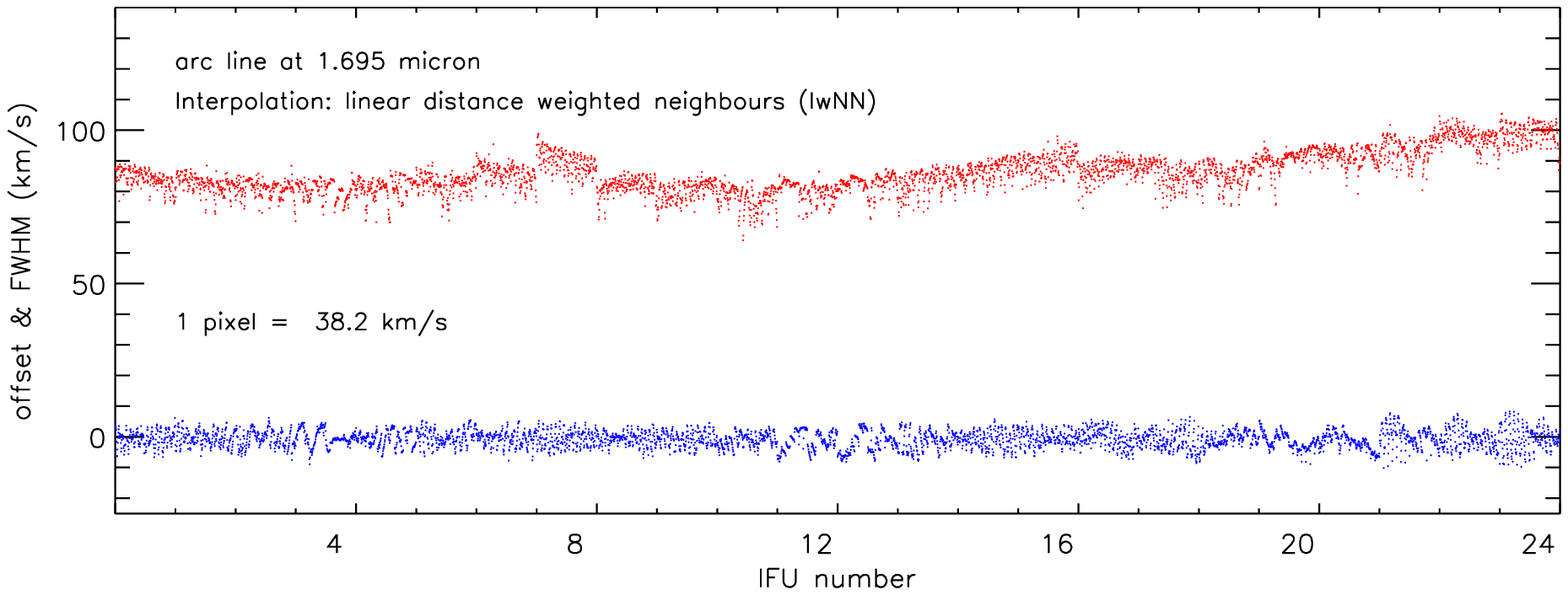}}
\resizebox{14cm}{!}
{\includegraphics{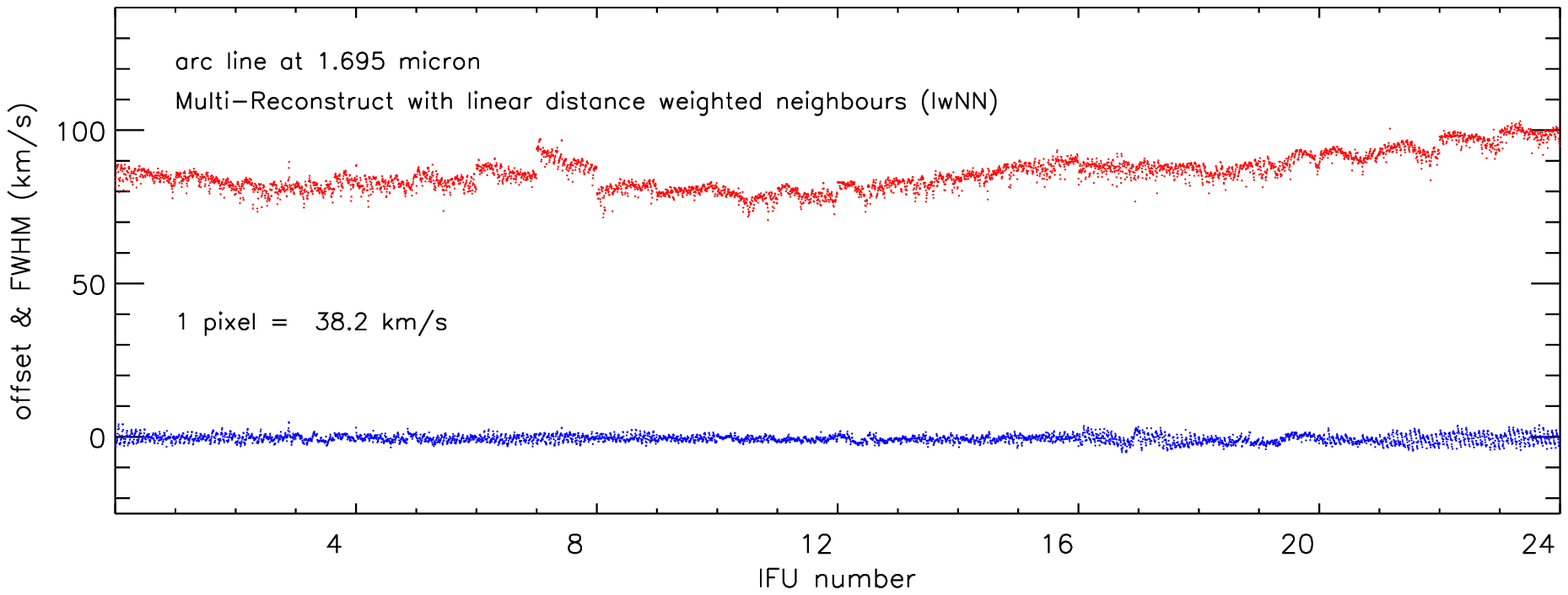}}
\end{center}
\caption{ \label{fig:lwnn} 
Spectral fidelity of an arc-line in the H-band as a function of spatial position across all 24 IFUs, when interpolating with a linearly distance weighted average of nearest neighbours.
The offset of the line from its nominal position (blue), and its FWHM (red), are shown.
Top: for a cube reconstructed from a single exposure.
Bottom: using the multi-reconstruct scheme described in Section~\ref{sec:multirecon} to generate a single cube from 6 input frames at different rotator angles.
The improvement can be quantified in terms of the rms error in the line position, which is 2.8\,km\,s$^{-1}$ for the single frame, and 1.2\,km\,s$^{-1}$ when using 6 frames simultaneously.
}
\end{figure*} 

This method makes use of all samples within a specified range -- typically 8 neighbours in 3 dimensions for the default range of 1.001 (in units of sampling distance).
The algorithm calculates a weighted average of these values, weighting each by the inverse of its distance from the grid point that is being interpolated.
It is a simple method with simple error propagation, which yields reasonable results as shown in the upper panel of Fig.~\ref{fig:lwnn}.
The spectral resolution is not as good as for the nearest neighbour because, in effect, the weighting scheme smooths the data to some extent; 
but the resolution is more consistent between different spatial locations.
There is also a big improvement in the line location.
The inclusion of several samples, and the smoothing effect, have worked to considerably reduce the spectral offset with respect to using just the nearest neighbour.
It should be kept in mind that for this method, and for all the weighted averages of neighbouring samples, increasing the search radius will lead to more smoothing (in 3D) of the data.

An alternative linear weighting scheme that was initially considered is kriging, which is commonly used in geological studies \citep{cla04}.
This has the advantage of being an optimal method, in the sense that the uncertainty of an interpolant is a minimum; and the weightings take into account the resolution of the data (or the spatial scales over which data are expected to change).
However, the comparisons of \cite{yan04} showed that, despite being a high quality method, it is also very slow, making it inappropriate for the pipeline.

\subsubsection{Quadratic Distance Weighting}

\begin{figure*}
\begin{center}
\resizebox{14cm}{!}
{\includegraphics{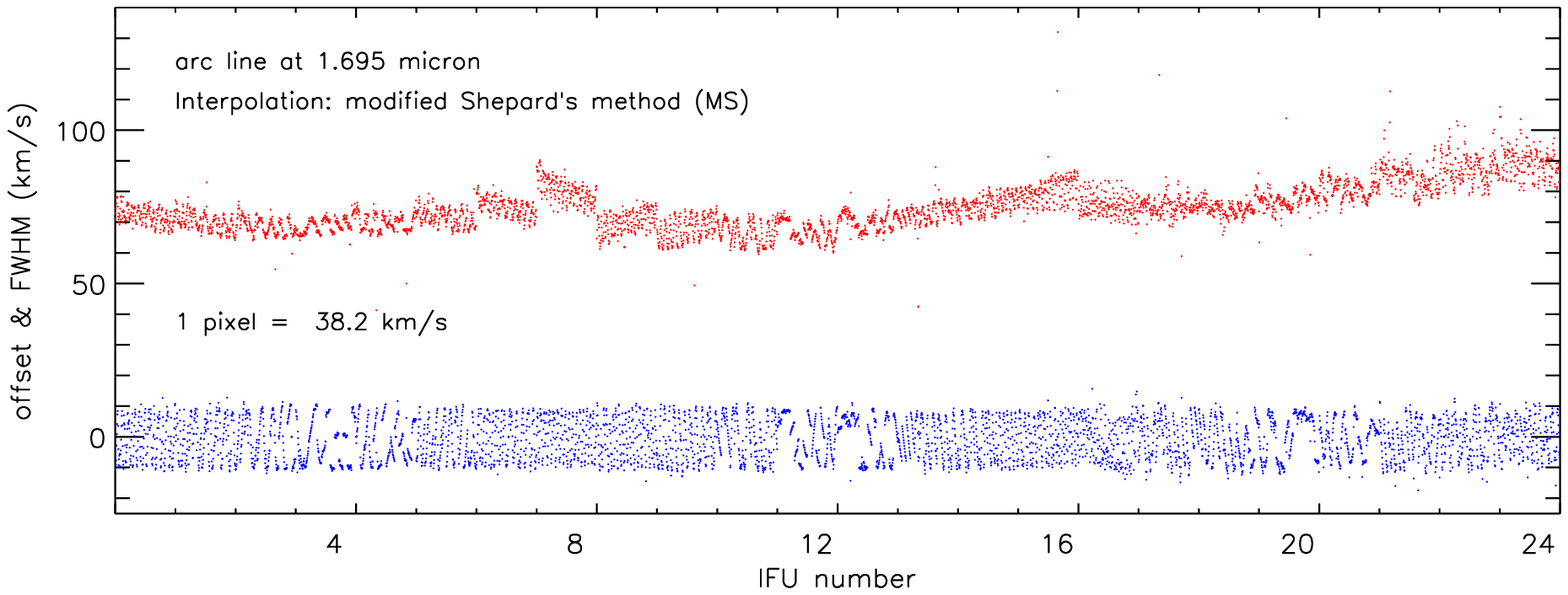}}
\resizebox{14cm}{!}
{\includegraphics{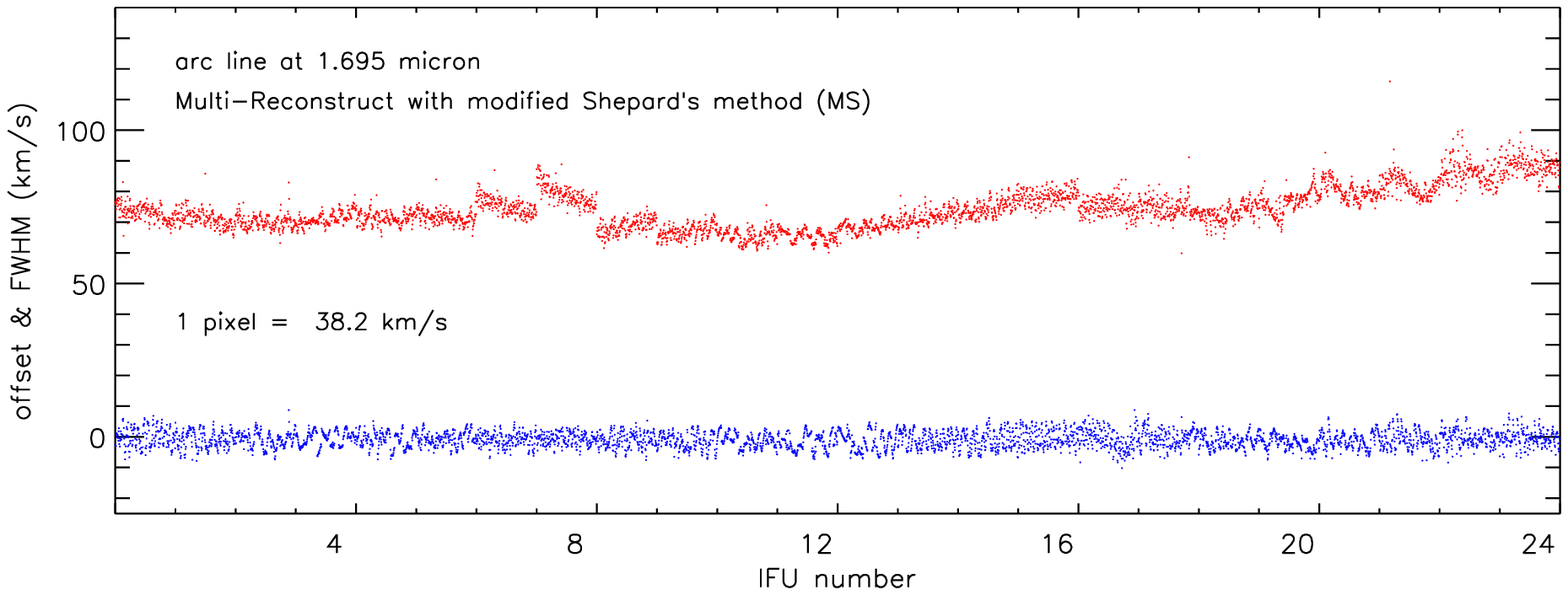}}
\end{center}
\caption{ \label{fig:ms} 
Spectral fidelity of an arc-line in the H-band as a function of spatial position across all 24 IFUs, when using the Modified Shepard's method for the interpolation.
The offset of the line from its nominal position (blue), and its FWHM (red), are shown.
Top: for a cube reconstructed from a single exposure.
Bottom: using the multi-reconstruct scheme described in Section~\ref{sec:multirecon} to generate a single cube from 6 input frames at different rotator angles.
The improvement can be quantified in terms of the rms error in the line position, which is 6.6\,km\,s$^{-1}$ for the single frame, and 2.7\,km\,s$^{-1}$ when using 6 frames simultaneously.
}
\end{figure*} 

Adjusting the scheme above to weight the samples by the inverse square of the distance (truncated at edge of the neighbourhood range) does not change the fidelity of the interpolation greatly.
Applying a greater weight to any sample(s) close to the required grid point means that, in effect, this method is intermediate between the nearest neighbour and the linearly distance weighted sum of neighbours.
And the fidelity is also intermediate.

An alternative quadratic scheme available in the pipeline is the modified Shepard's method in which the relative weights decrease to zero at the truncation radius.
It is described by \cite{ren88}, who also addresses the issue of extending the method to 3 dimensions.
The comparisons performed by \cite{yan04} suggest this is an accurate and relatively fast method, at least for surface fitting.
Our implementation differs from this method by not fitting a (multi-variate) quadratic function to the samples around the interpolant grid point, because the distribution of radii (and also typically the level of noise), is insufficient to yield a reliable fit.
The fidelity of the interpolation, shown in the upper panel of Fig.~\ref{fig:ms}, is again intermediate between linear distance weighting and nearest neighbour: the spectral resolution of an isolated line is improved, but its position shows a larger variation.

\subsubsection{Cubic Spline}

\begin{figure*}
\begin{center}
\resizebox{14cm}{!}
{\includegraphics{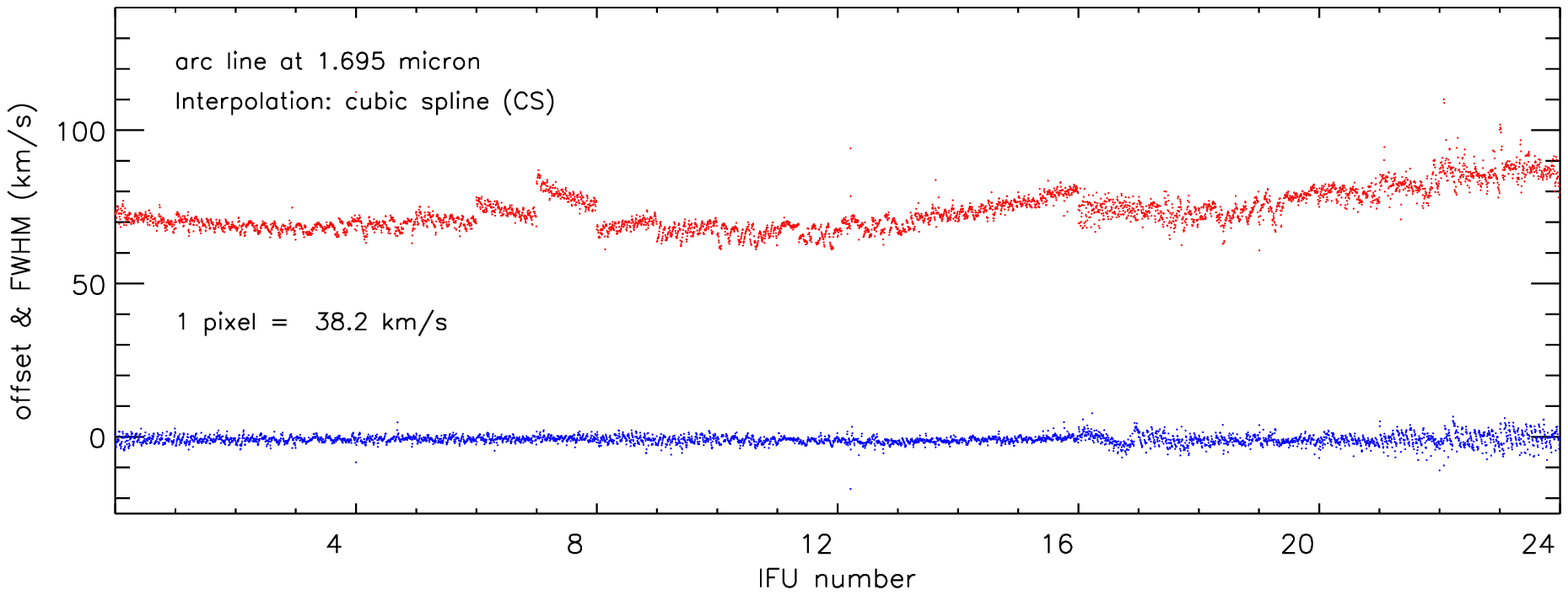}}
\end{center}
\caption{ \label{fig:cs} 
Spectral fidelity of an arc-line in the H-band as a function of spatial position across all 24 IFUs, when using the cubic spline method for interpolation (with a single exposure).
Note that this is, in fact, a series of 1-dimensional interpolations
The offset of the line from its nominal position (blue), and its FWHM (red), are shown.
For comparison to the other interpolation methods, the rms error in the line position, which is 1.5\,km\,s$^{-1}$.
}
\end{figure*} 

This is the default method used in the pipeline, for reasons which are clear from the fidelity of the interpolation shown in Fig.~\ref{fig:cs}.
It is a series of two or three 1-dimensional interpolations, which make use of the fact that all the samples (detector pixels) within any given slitlet lie along a straight line when projected onto the sky, and that the spacing across the slitlets is fixed and uniform.
Bicubic (or even tricubic; \citealt{lek05}) spline interpolation cannot be applied to KMOS data because these methods require a regular 2D (3D) grid of samples.
As such, a cubic spline is first performed along each slitlet.
No interpolation is needed across the slitlets, unless the output pixel scale is set to be different from the 0.2\arcsec\ default, in which case a second cubic spline is performed.
The final interpolation is along the spectral axis.
But for spline interpolation, the limited sampling with respect to the resolution in the spectral direction would lead to severe problems with overshoot at the base of strong emission lines.
To overcome this, rather than tensioning the spline which reduces the interpolation order to linear, we have implemented polynomial interpolation along that axis.
This method gives the best results in terms of fidelity: it maintains the native resolution of the data, and achieves good precision in line location.

\subsection{Multi-Reconstruct}
\label{sec:multirecon}

\begin{figure}
\begin{center}
\resizebox{\hsize}{!}
{\includegraphics{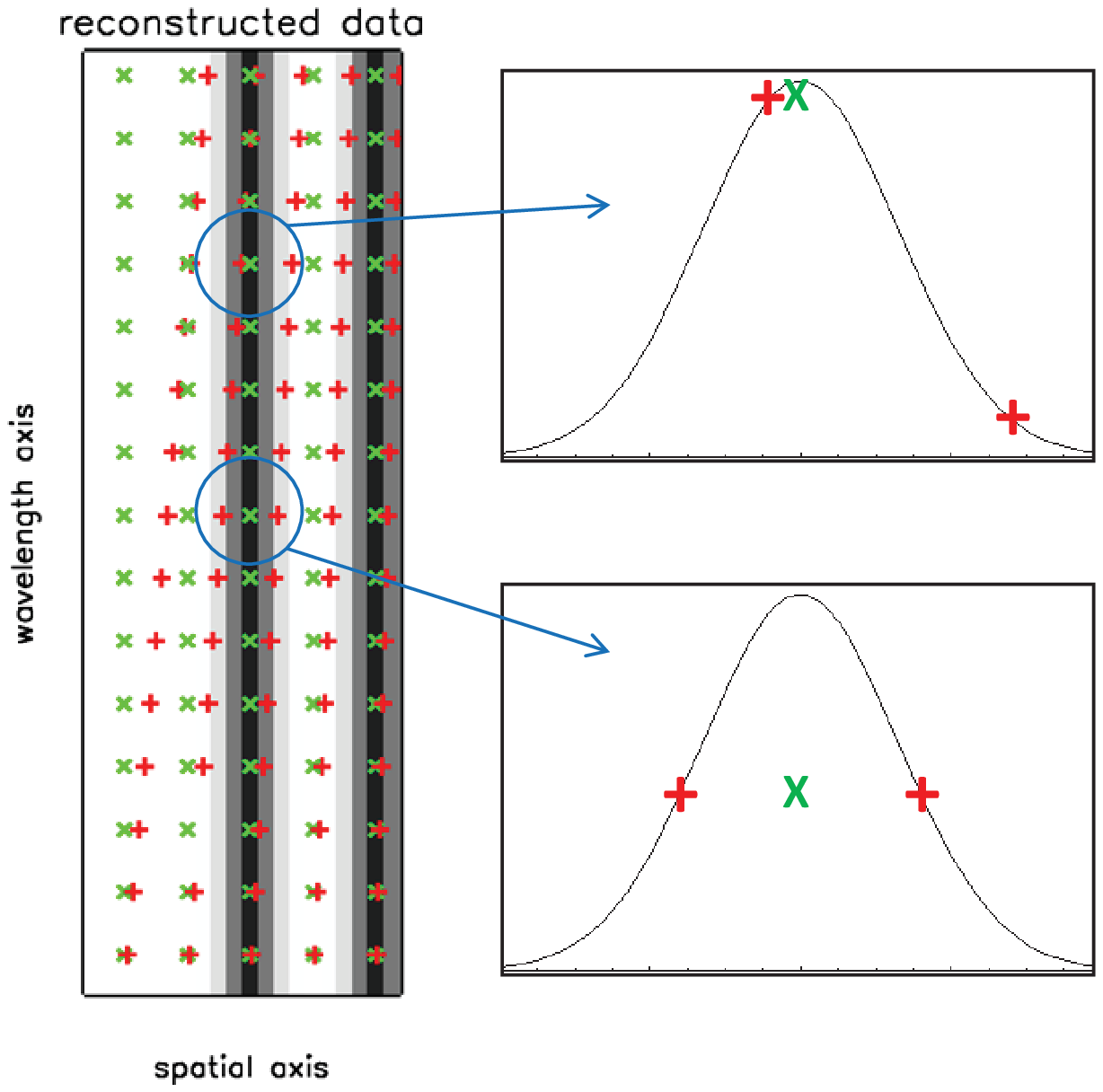}}
\end{center}
\caption{ \label{fig:sampling} 
Illustration of how the performance of the 3D interpolation schemes are limited by sampling:
because these all use weighted sums of neighbouring points, the interpolant cannot be greater than the largest neighbour.
The left panel (panel (d) of Fig.~\ref{fig:interp}) shows sampled points (red) and the interpolation grid (green).
The right panels show a horizontal cut through 2 different regions.
Lower: the grid point to be interpolated lies midway between two sample points, and so cannot trace the true peak of the profile.
Upper: since one of the samples is almost at the peak of the profile, the interpolant can provide a more precise estimate of the true value of the peak.
It is straightforward to realise that this means the spectral shape of a spatially compact source (e.g. a star) will have a slow ripple pattern at a frequency corresponding to how often the dispersion axis crosses a column of pixels on the detector, as shown in Fig.~\ref{fig:wiggle}.
The multi-reconstruct recipe overcomes this limitation.
}
\end{figure} 

\begin{figure}
\begin{center}
\resizebox{\hsize}{!}
{\includegraphics{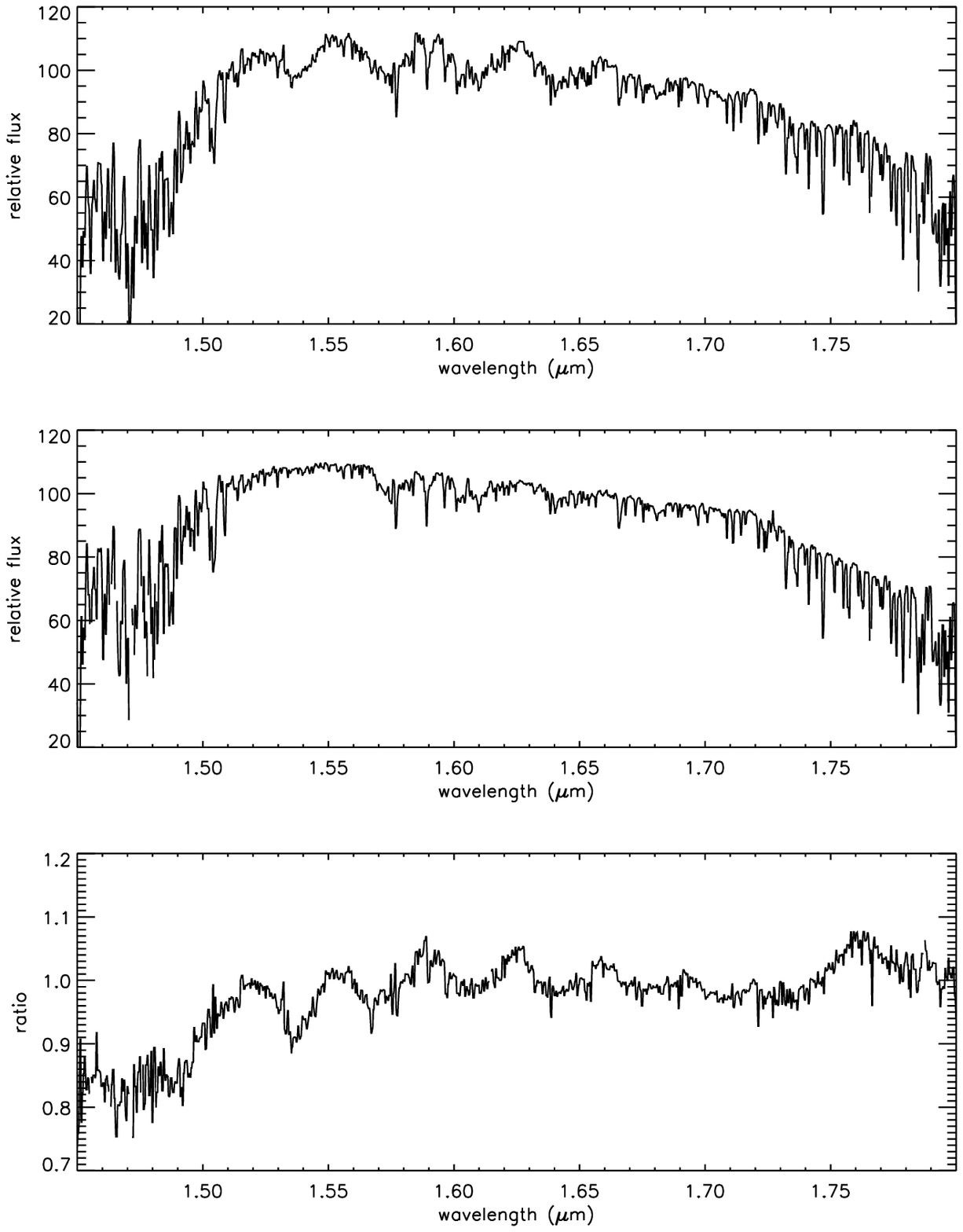}}
\end{center}
\caption{ \label{fig:wiggle} 
Top panel: Rather extreme example of the ripples, with a period of 150 spectral pixels or about 0.03\,$\mu$m, that can be seen in the spectra of individual spatial pixels if the seeing is so good that the star is spatially undersampled.
Middle panel: spatially integrated spectrum extracted from a different IFU which does not show the ripple effect.
Bottom panel: ratio of the 2 spectra to show the ripples more clearly.
This example is for cubic spline interpolation in 0.35\arcsec\ seeing (as measured in the H-band on the star itself).
For nearest neighbour reconstruction, the effect is more severe and can be manifested as saw-tooth discontinuities.
In either case, it can only be avoided by better sampling, which is one of the things provided by the multi-reconstruct recipe (described in Sec.~\ref{sec:multirecon}).
}
\end{figure} 

A rationale for the calibration concept described in Section~\ref{sec:mapdistort} is to combine multiple exposures during the reconstruction itself, rather than afterwards.
This is the purpose of the multi-reconstruct recipe (which was initially planned just as an alternative reconstruction method, but out of necessity has evolved into a full science reduction in its own right).
The recipe simply puts all the data and associated calibrations into meta-frames, and then treats these meta-frames as it would an individual exposure.
Doing this provides two main advantages, discussed below.

Reconstructing each cube separately means that they have to be shifted to a common grid in the World Coordinate System (WCS) before they can be combined.
If, during the observations, the dithering is performed in sub-pixel steps, this process will then require an additional interpolation.
That can in fact be avoided by using dithers that are multiples of half a nominal pixel (i.e. 0.1\arcsec) and reconstructing the data with a  0.1\arcsec\ grid spacing.
There will then only be integer shifts between the cubes, and so no additional interpolations are required.
More simply, the issue is avoided completely when using multi-reconstruct. 

The second advantage is that it overcomes the sampling limitation of all the 3D interpolation schemes, which is illustrated in Fig.~\ref{fig:sampling}.
This problem arises because these schemes are all based on weighted averages of neighbouring sample points, and so the interpolant cannot be greater than the largest of the those.
For any source that is compact along at least one axis, and very extended along another (e.g. a star is spatially compact and spectrally extended), there will be low frequency, but possibly severe, ripples along the extended axis.
An example of this effect can be seen in Fig.~\ref{fig:wiggle} which shows the H-band spectrum from an individual spatial pixel of a star observed with 0.35\arcsec\ seeing (as measured in the H-band on the star itself).
These ripples can only be avoided by finer sampling.
Combining multiple exposures at various dither positions during the reconstruction will provide exactly that, as demonstrated in the lower panels of Figs.~\ref{fig:nn}--\ref{fig:ms}.
Here, the flexure within KMOS (see Section~\ref{sec:flex}) has been used to illustrate quantitatively the potential gain offered by the kmo\_multi\_reconstruct recipe.
We have reconstructed a single set of 24 arc-lamp cubes using, as simultaneous input to the recipe, exposures at all 6 different rotator angles (from 0--300$^\circ$ at 60$^\circ$ intervals).
The resulting spectral fidelity, as measured by the rms error in the measured wavelength of an emission line, is improved by a factor of 2--2.5.
Based on this simple example, it seems reasonable to expect significant gains in both spatial and spectral quality for science observations taken at a variety of dither positions and over a range of rotator angles.

We note that cubic spline interpolation is not ideally suited to the multi-reconstruct method.
The main reason is a combination of the fact that, by definition, the interpolated function between a set of points goes through those points, and that astronomical data is typically noisy.
When working with several exposures simultaneously, one may often find situations where there is a sequence of samples close together with different values, which will inevitably lead to high frequency oscillations in the interpolated function.
This, in turn would amplify the noise in the resampled data considerably, a problem that can only be overcome with a smoothing or approximating spline function \citep{rei67,boo01}.

\subsection{Mosaics}

When executing an observation with KMOS, there is a distinct mode that can be used for mapping a large contiguous area.
But from the data processing perspective, this is just a specific instance of shifting and combining exposures.
Instead of having small dithers in order to reach greater depth on a restricted field, the dither size is increased until it is nearly the size of the IFU field of view.
In this limiting case, the total area covered is large while the overlapping regions are small.
But the pipeline software is the same in both cases.
The only difference is how it handles the edges of each IFU field when performing sub-pixel shifts to align the data to a common WCS grid.
The default for dithering is that no extrapolation is performed,
In this case, one will usually lose the row or column of pixels along the edge in the direction the data are shifted, and so the shifted field is slightly smaller than the original.
But when putting together a mosaic, this would lead to a `wire frame' of empty pixels around the individual IFU fields.
So for mosaics, extrapolation is enabled.
The multi-reconstruct approach avoids this issue since, when taken together, the combined sampling of all the exposures only stops at the edge of the mosaic, rather than at the edge of each individual IFU.

When putting together a mosaic, temporal changes in the sky background and also in atmospheric throughput can lead to discontinuities in the flux levels between the various pieces and other edge effects.
With 72 individual pieces for the small mosaic (9 pointings, each with 8 IFUs, covering a total of $32\arcsec \times 16\arcsec$), 384 pieces for the large mosaic (16 pointings, each with 24 IFUs, covering a total of $65\arcsec \times 43\arcsec$), or more than 1000 pieces if the same mosaic is observed several times with small offsets, compensating this can be a very complex and computationally intensive task.
The way in which one might deal with it depends on the type of source one is looking at (e.g. a sparse distribution of stars or continuous extended emission from a galaxy) as well as on the science goals.
Because of this, the pipeline itself makes no attempt to address this problem.

%

\section{Flexure}
\label{sec:flex}

\begin{figure}
\begin{center}
\resizebox{\hsize}{!}
{\includegraphics{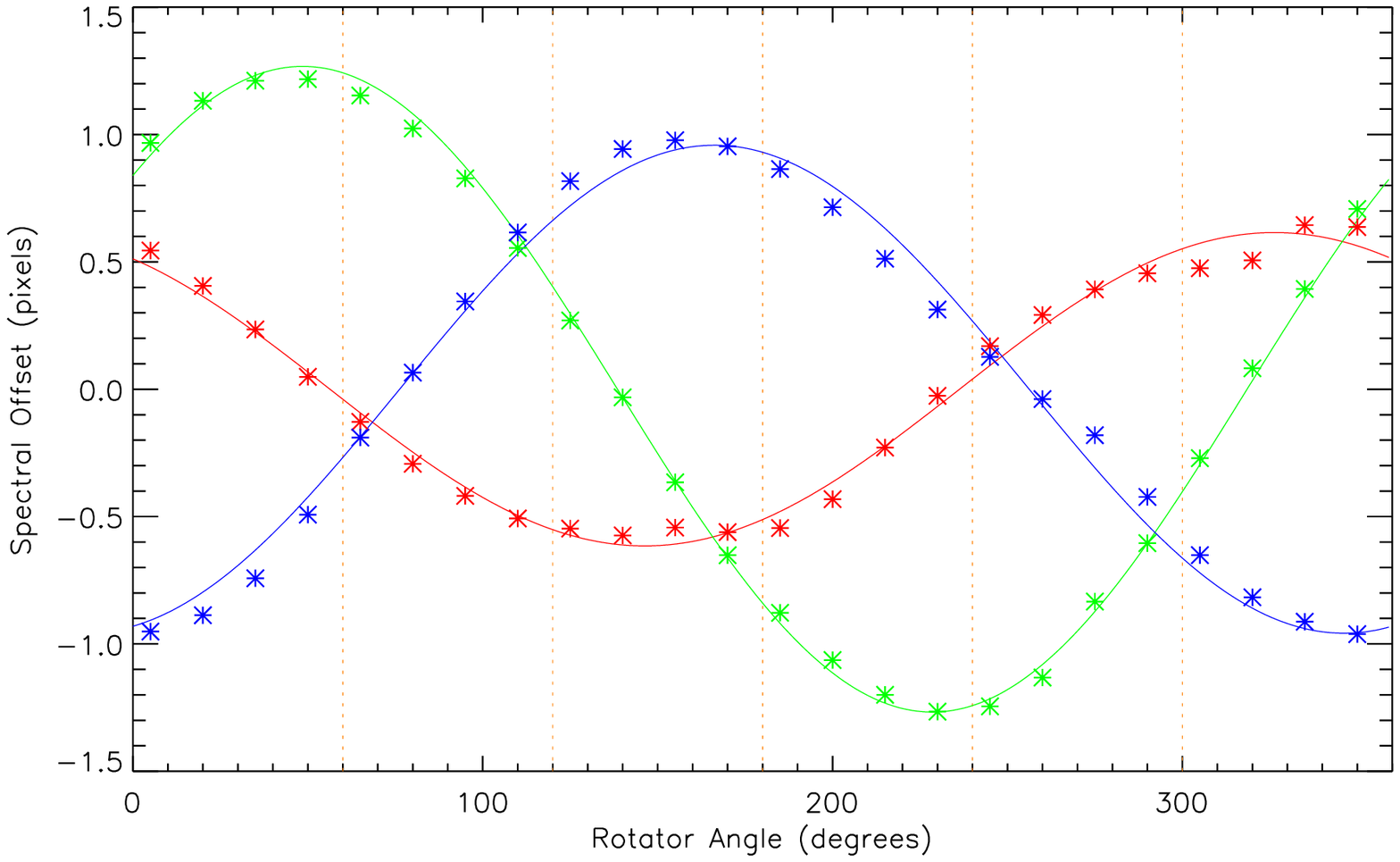}}
\resizebox{\hsize}{!}
{\includegraphics{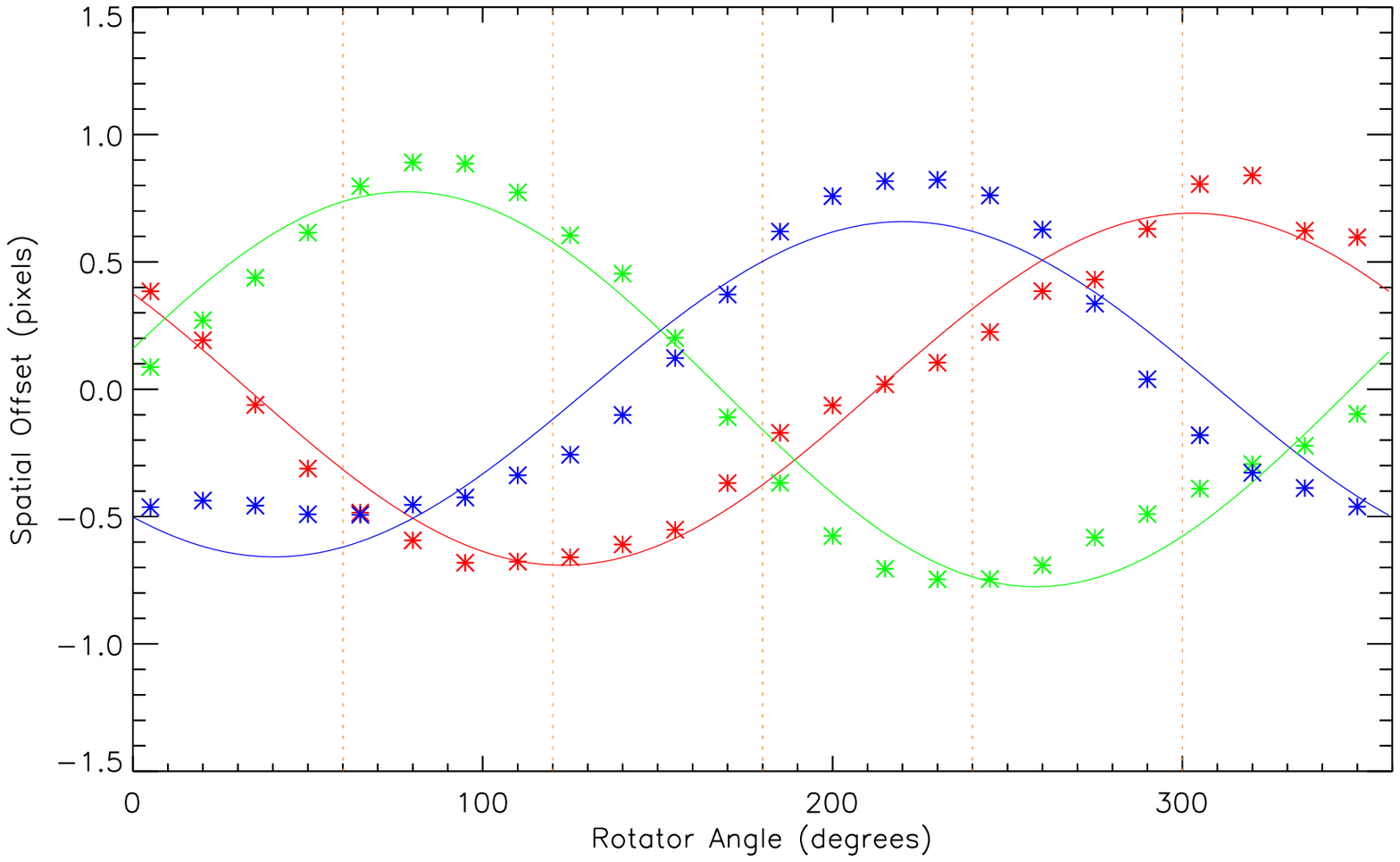}}
\end{center}
\caption{ \label{fig:flex} 
Spectral (top) and spatial (bottom) flexure measured from the location of an arcline (vertical shifts) and slitlet edges (horizontal shifts) respectively on raw detector frames, as a function of rotator angle.
The 3 colours represent the 3 instrument segments (red/green/blue for segments 1/2/3).
The lines trace sinusoidal fits to the data: while this is a good match for the spectral flexure, it is only an approximation to spatial flexure.
Importantly, in both cases, the scale of the shift is $\sim\pm$1\,pixel, indicating that which pixels on the detector are illuminated depends on rotator angle.
}
\end{figure} 

There are several sources of flexure, or other causes of non-repeatability, within KMOS. 
Most are the result of the instrument being mounted at the Nasmyth focus and rotating around a horizontal axis so that the gravity vector on the 3-tonne instrument changes.
In decreasing order of severity and impact on data processing these are:
\begin{enumerate}[(i)]
\item
{\it Spatial and spectral flexure}. This is likely to arise within the spectrographs because it causes a shift of the data on the detector, both horizontally and vertically (Fig.~\ref{fig:flex}). The shifts approximately follow a sinusoidal pattern, for which the phase differs between instrument segments. Its amplitude is $\sim\pm$1\,pixels on the detector. The way in which this flexure is addressed by the pipeline is described below in Section~\ref{sec:compensate}.
\item
{\it Global spatial flexure}. This has no impact on the location of the data on the detector but instead leads to a rotator dependent circular shift of the IFU pointings on sky (all in the same sense) with an amplitude of $\sim$0.3\arcsec. As such, it is probably due to the whole instrument sagging slightly at the rotator interface, and matches the predictions from the finite element analysis of the instrument design. In order to track the shifts, one IFU needs to observe a reference source -- which can also be a science target, as long as it is bright enough for centering in a single exposure.
\item
{\it Grating precision}. This is not flexure, but a non-repeatability that appears to be related to the precision of the grating position when the wheel is moved between bands and back again (although that has not been investigated exhaustively). It is manifested as a shift of the data on the detector along the slitlets with an rms of $\sim$0.2\,pixels. The effect on the reconstructed data is to make one edge of the field (along the ends of the slitlets) brighter and the opposite edge fainter. Because it is random, it can only be addressed by re-measuring the slitlet edges after a grating change. This is done by the pipeline for the illumination correction with the sky-flat data. 
\item
{\it Temperature dependency}. Only the detectors are temperature stabilised, while the temperature of the opto-mechanics inside the crysotat may vary within the range of approximately 114-122\,K, possibly related to changes in the ambient temperature. This leads to a stretching of the data on the detector, with an amplitude of about 0.1\,pixels across each detector for every 1\,K change in temperature (see Fig.~\ref{fig:temp_shift}). Since the rate of temperature change is not more than 1\,K per day, the only requirement arising is that calibrations are taken within 1 day of the observations (as is standard procedure) so that the differential stretching remains at a negligible level.
\item
{\it Differential arm flexure}. This is very small, and as yet there are no convincing measurements of it in the data. 
There are hints for it in some datasets that would indicate it is at a level of $\sim$0.03\arcsec, which is at the limit of what can be reliably measured.
\end{enumerate}

\subsection{Compensating Spatial and Spectral Flexure}
\label{sec:compensate}

\begin{table*}
\caption{Summary of recipes available in SPARK for processing KMOS data}
\label{tab:recipes}
\begin{tabular}{ll}
\hline\hline
recipe & description \\
\hline
\multicolumn{2}{l}{\hspace{1cm}Basic calibration and pipeline recipes} \\
kmo\_dark & Create master dark frame \& preliminary bad pixel mask \\
kmo\_flat & Create master flatfield frame, spatial calibration frames, and final badpixel map \\
kmo\_wave\_cal & Create a spectral calibration frame \\
kmo\_illumination & Create calibration frame to correct spatial non-uniformity of flatfield \\
kmo\_std\_star & Create telluric correction spectra and calculate zeropoint \\
kmo\_sci\_red & Perform standard science reduction on a set of observations \\
kmo\_multi\_reconstruct & Reconstruct and combine cubes simultaneously from multiple frames \\
\multicolumn{2}{l}{\hspace{1cm}Miscellaneous Tools} \\
kmo\_arithmetic & Perform basic arithmetic on frames and cubes \\
kmo\_sky\_tweak & Subtract sky cubes from object cubes, applying scaling based on OH line strengths \\
kmo\_make\_image & Collapse a cube spectrally to create a spatial image \\
kmo\_extract\_spec & Extract a spectrum from a cube \\
kmo\_fit\_profile & Fit spectral or spatial profiles to lines or objects \\
kmo\_stats & Calculate basic statistical properties of a frame or cube \\
kmo\_reconstruct & Perform cube reconstruction on a single frame\\
kmo\_combine & Combine reconstructed cubes \\
kmo\_shift & Shift a cube spatially by a sub-pixel increment \\
kmo\_rotate & Rotate a cube spatially \\
kmo\_noise\_map & Generate a noise map from a raw frame \\
kmo\_copy & Copy a section of a cube \\
kmo\_fits\_check & Check contents of a KMOS fits file \\
kmo\_sky\_mask & Create a mask of spatial pixels that can be considered as sky \\
kmo\_fits\_strip & Strip extensions (noise, rotator angles, and/or empty extensions) from a file \\
\hline
\end{tabular}
\end{table*}

The amplitude of the flexure shown in Fig.~\ref{fig:flex} indicates that it cannot be compensated simply by modifying the spatial/spectral identification associated with the same set of detector pixels. 
Instead, it is absolutely necessary to take calibrations at different rotator angles, because the set of pixels that are illuminated changes as the instrument rotates.
The standard procedure is to take calibrations every 60$^\circ$ around a full circle, providing a set of 6 flatfield and arc frames.
In addition, a more specific set of rotator angles are also used, that are matched to the observations taken during the previous night.
The disadvantages of doing this are that it is time consuming, it significantly increases the size of the calibration products, and it makes calibrating and handling the science data significantly more complex.
On the other hand, these multiple calibration frames have no impact on the algorithms used for the data processing, are largely transparent to the user, and make a reasonable first-order correction to the most serious flexure.

The pipeline provides additional corrections to compensate for the gap between rotator angles at which calibrations are taken.
Spatially, this involves linearly interpolating between each of the spatial calibration frames adjacent to the data being reconstructed.
This is actually straightforward since, to a precision of a few milliarcsec, there is only a constant offset between any pair of such frames for each set of four IFUs (half a segment).
This process means that one might lose some data at the edges of the slitlets, depending on how much the edge locations differ between the two angles being used.
However, tests on the commissioning data show it significantly improves the spatial reliability of the reconstructed field.
And, because the change is made to the recipe's internal copy of the calibration files before they are used for the reconstruction, it preserves the pipeline concept of having only one step where interpolation is performed on science data.

Spectrally, the correction is measured using the atmospheric OH lines, which is easily done in most science exposures.
This uses the method described by \cite{dav07} but, as above, applies the correction to the recipe's internal copy of the wavelength calibration file.
Doing so requires a double pass of the data.
In the first pass, a cube is reconstructed so that the wavelength offset can be measured easily.
This offset is then applied to calibration, and the data reconstructed again.
Thus, the final science product has only undergone a single interpolation step.

%

\section{Work Flow}
\label{sec:workflow}

A description of the design of the pipeline is given in \cite{agu13}.
This includes a full explanation of each recipe, together with a flow chart, list of input frames and parameters, and a list of output products and quality control parameters.
In addition, it provides a definition for the formats of raw and processed KMOS data.
There is also a separate guide about using the pipeline and its recipes \citep{dav13}.
The reader is referred to those documents for details of the pipeline and its usage.
In this section, we provide only a short overview of the standard steps involved in reducing KMOS data, and focus instead on the key issues that will be encountered.

\subsection{Standard Processing Steps}

In many cases, science data can be processed using a single recipe kmo\_sci\_red.
If a user prefers not to do that, all the functionality of this recipe can be performed using other recipes provided as tools, although the task is made more complex by the multiple rotator angles used for calibration, and the mixing of object and sky data in individual exposures which is a natural result of having 24 IFUs. 
The kmo\_sci\_red recipe deals with both of these issues automatically.
It takes as input a set of observations together with a set of calibration products (at multiple angles) in the same waveband.
Selecting the calibration angle that best matches (i.e. is closest to) the data is straightforward.
To enable sky subtraction, the recipe first identifies, for each IFU, whether each exposure contains object or sky data.
It then works through the IFUs in sequence, considering each time only data associated with that specific IFU.
For each object frame, it then allocates a sky frame that is taken closest in time (sky frames may be used twice, depending on how the observing sequence was defined -- this is acceptable if there are spatial dithers between the object frames for which it is used).
Thus, a table is generated that provides a matching between object and sky exposures for each IFU independently.
While this will often match what the observer intended, there are many other ways to match up object and sky data, perhaps even between different IFUs of the same exposure -- although in doing this one would need to take care of the differing spectral line profile across the instrument.
A simple method to achieve this will be included in a future release of the pipeline (see Section~\ref{sec:conc}).

Once the appropriate calibration angle has been selected, and the object-sky pairings assigned, 
the recipe can proceed with the normal steps.
Because each IFU in each exposure may have to be treated differently right from the start, the pipeline works with a single IFU at a time: subtract the sky frame, divide by the flatfield, reconstruct the cubes (applying the flexure corrections described in Section~\ref{sec:compensate}), divide out the telluric spectrum, apply an illumination correction, align the cubes, and finally combine them.
The result is a set of files which contain the combined datacube, and an associated noise cube, for each IFU.

Several specific options are worth discussing because they address the problems related to the varying background.
The first is to apply the spectral scaling that compensates for the variable OH airglow line fluxes, which is done exactly as described in \cite{dav07}.
In the KMOS pipeline, this option is independent of the wavelength flexure correction; but will only work well if that correction is performed.

The second option is to derive and apply a constant offset correction to match (i.e. set to zero) the residual background levels between exposures -- which has a crucial impact on flux conservation.
Although typically small, without such a correction the combined image can suffer from discrete jumps in the background level at the locations of the edges of individual cubes.
This has a highly detrimental impact on observations of faint targets, and so has to be corrected.
The small field of view of each IFU means that using the straight median of the pixels is likely to yield an overestimate of the background, which can be especially severe for extended continuum source (although less so for purely emission line sources).
In order to obtain an estimate of the background that does not suffer from too much bias in either direction, we settled for the mode of the pixel values after excluding the brightest 25\%.
The user has to decide whether to apply this correction: it is valid if there are sufficient regions of sky in an IFU field, but may be completely wrong if the IFU field is filled with (part of) an extended object.

Flux conservation is also awkward, the reason being that the first step in data reduction is to subtract a sky frame; 
and because the sky varies, a correction for the residual background needs to be made.
If this is not done, a slightly negative residual background might match the positive counts in a faint target, so that the total flux in the cube is nearly zero.
In extreme cases, this can cause a flip in the sign of the entire data set.
Thus a robust and precise estimation of the background level is crucial if the flux conservation is to be applied correctly.
As a result, this too is left to the user to decide whether they wish to apply it.

Finally, an unexpected issue noticed during commissioning was that the mean of the pixel values in any given readout channel on the detectors varies in time.
The cause of this is not clear.
Although only on the scale of about $\pm$2\,counts, it can have a serious impact on data of very faint objects, producing a striping effect across the reconstructed and sky-subtracted IFU fields of view.
At the current time, it is expected that changes to the detector control electronics will moderate, or fully remove, this effect.

\subsection{Error Propagation}

Error propagation is an important part of any data pipeline.
But a user should always be wary of how noise estimatation is implemented, since it is often simplified, and it is difficult to include systematics which can become very important.
\cite{vac04} provide a detailed and generally applicable analysis of the variance for pixel values in frames read out from near-infrared detectors, but this also does not take into account systematics effects that arise between different frames.
Instead, the best and most reliable noise estimate is made during the final stage of data reduction, when frames (cubes) are combined -- as long as at least several frames are combined.
Then one can directly assess the distribution of values at each location in the final data set, and use that to derive the noise in the output data directly, thus including any systematics and bypassing approximations made during error propagation.

But even after the data are fully reduced, when analysing the data, one still needs to be cautious about uncertainties.
For example, when extracting integrated spectra from SINFONI \citep{eis03}, standard error propagation yields an underestimate of the true noise.
This is because neighbouring pixels are partially correlated, and so the noise increases faster with aperture linear size $N$ (where $N = \sqrt{A}$, for an aperture of area $A$) than the Gaussian scaling $N\sigma$ \citep{for09}.
These authors found that a function of the form
\[ \sigma_{real} \, / \, [N \, \sigma_{pixel}] = a \, + \, b \, \rm{log}(N) \]
provides a good description of the actual noise behaviour when extracting integrated spectra, and that the noise in the spectra was about a factor 2 higher than the simplistic expectation.
Inevitably, including this level of complexity in the pipeline is not practicable given that the constants $a$ and $b$ must be derived separately for each dataset (since they will depend on number of frames, sky subtraction strategy, dither sizes, exposure time, etc.); 
and how they are derived (or even if they can be derived) will depend on the sources in the IFU field.

The KMOS pipeline includes simple error propagation.
The initial noise estimate is derived for all valid pixels in each raw frame based on the effective read noise (which depends on exposure time) and gain.
When a mathematical operation is performed on frames or pixels, the associated noise frame is updated appropriately.
The last stage, when individual frames are combined to produce the final output cube, is the most important.
Here, the pipeline does attempt to include systematics.
If, at any given 3D location in the output cube, and after the rejection iterations, at least 3 values from the input cubes remain to be combined, the noise estimate is based on the standard error of those values.
At other locations where this criterion is not met, only the more simple propagated approximation to the noise can be used.

\subsection{Cosmetics}

The method used to identify bad (and unused) pixels is not sensitive to pixels that are intermittently or only temporarily affected.
These will remain in the data unless a sufficient number of dithered cubes are combined at the end of the processing.
That is not always the case, and so it can be helpful to cosmetically clean the reconstructed cubes.
One of the most successful routines for cleaning images is L.A.Cosmic, developed by \cite{vandok01}.
This is difficult to apply to the 2D (detector based) KMOS data because of the numerous OH lines in the data.
Since in most cases these are only barely Nyquist sampled, the routine is likely to identify them as bad pixels and remove them.
Instead, we have developed a version of this routine that works with a 3D kernel on the reconstructed data cubes (and is available independently of the pipeline).
While working in 3D adds little to the success of identifying and replacing bad pixels (both positive and negative), the routine includes stages to first account for structure in the cube such as stars or OH lines.
Numerous tests on both SINFONI and KMOS data have shown that this measure appears to be robust, and that applying the routine to reconstructed cubes is both safe and effective.

%

\section{Examples}
\label{sec:examples}

\begin{figure}
\begin{center}
\resizebox{\hsize}{!}
{\includegraphics{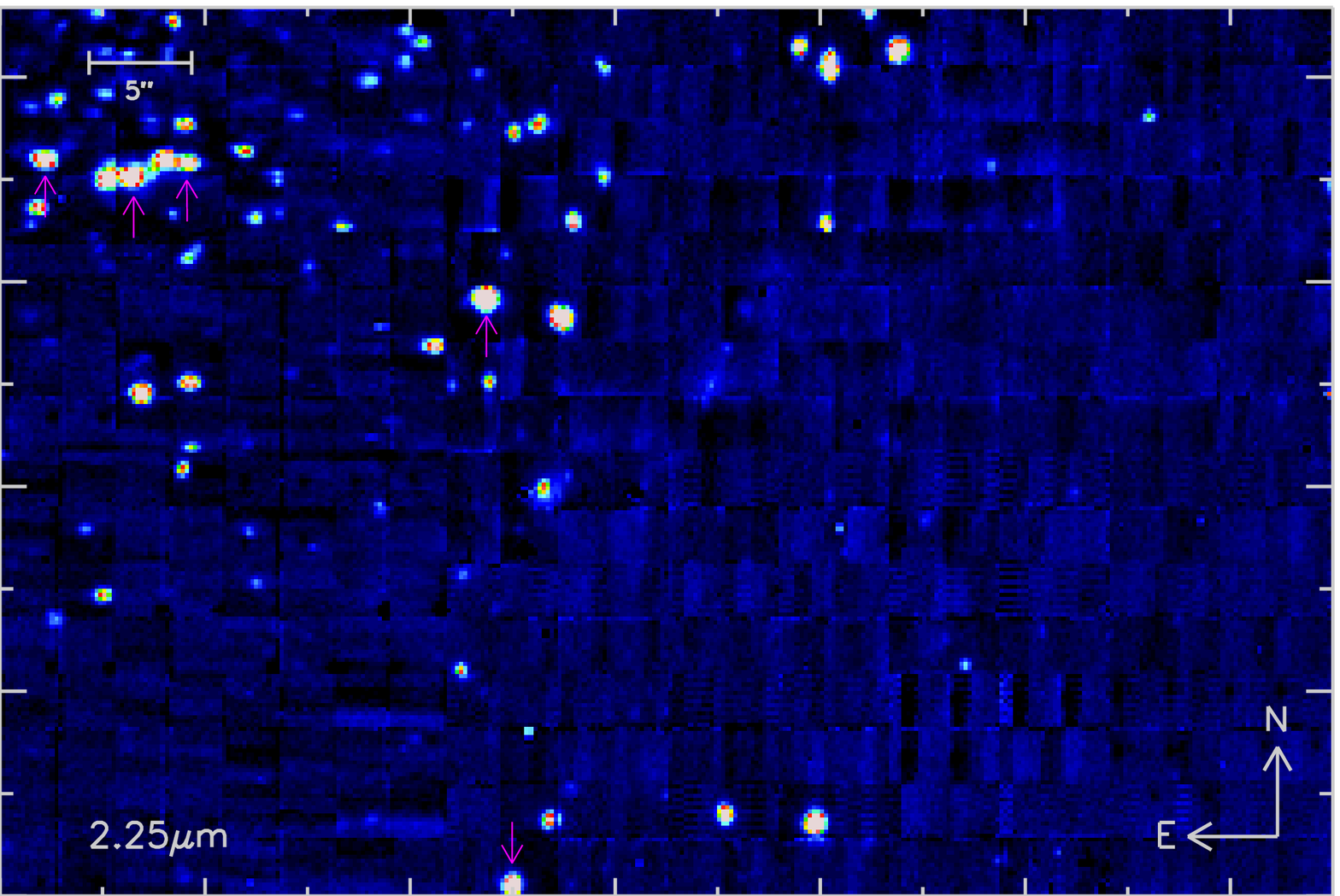}}
\resizebox{\hsize}{!}
{\includegraphics{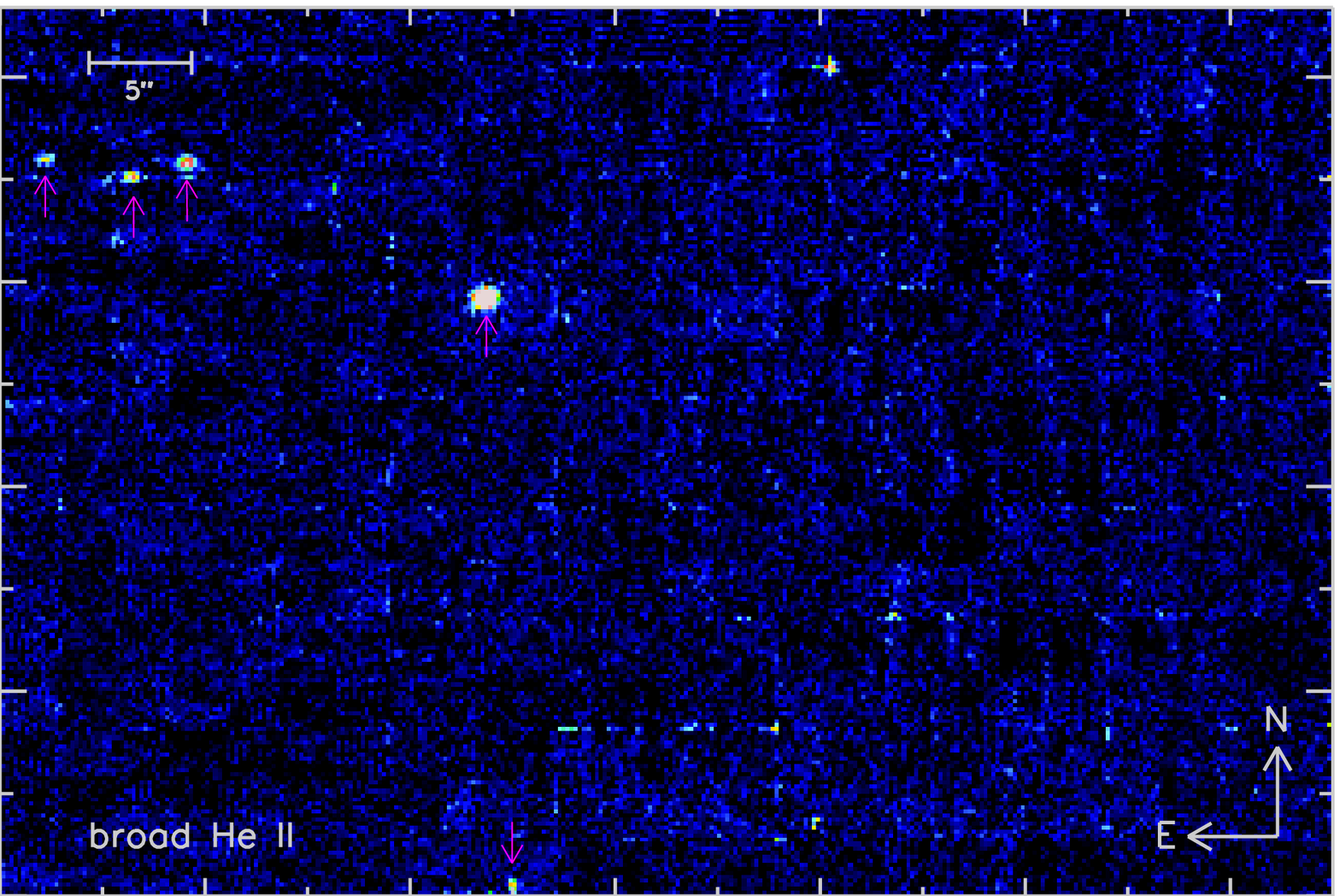}}
\resizebox{\hsize}{!}
{\includegraphics{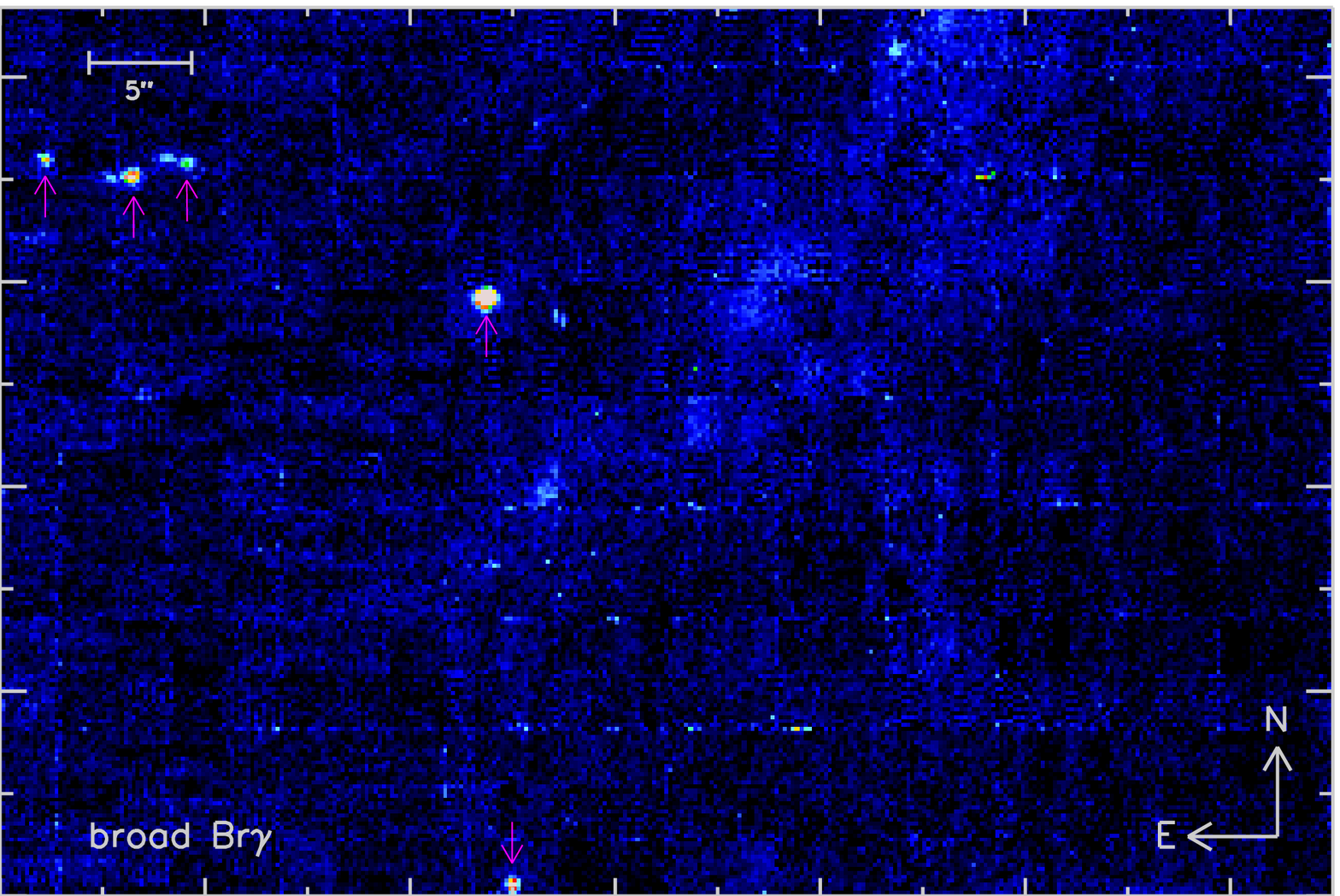}}
\resizebox{\hsize}{!}
{\includegraphics{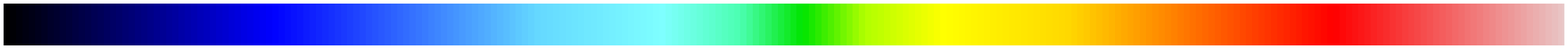}}
\end{center}
\caption{ \label{fig:r136im} 
Images from a KMOS mosaic of a region just to the south of the core of R\,136.
Top: 2.2--2.3\,$\mu$m continuum.
Middle: broad He\,II line emission at 2.19\,$\mu$m.
Bottom: broad Br$\gamma$ line emission (the faint extended structure is narrow line emission that has not been fully rejected).
The Wolf-Rayet stars, for which spectra are drawn in Fig.~\ref{fig:r136wr}, have been identified in each plot by magenta arrows.
All panels have been drawn on a linear scale using the colour bar shown underneath, with lowest values in black/blue and the highest values in red/white.
}
\end{figure} 
\begin{figure}
\begin{center}
\resizebox{\hsize}{!}
{\includegraphics{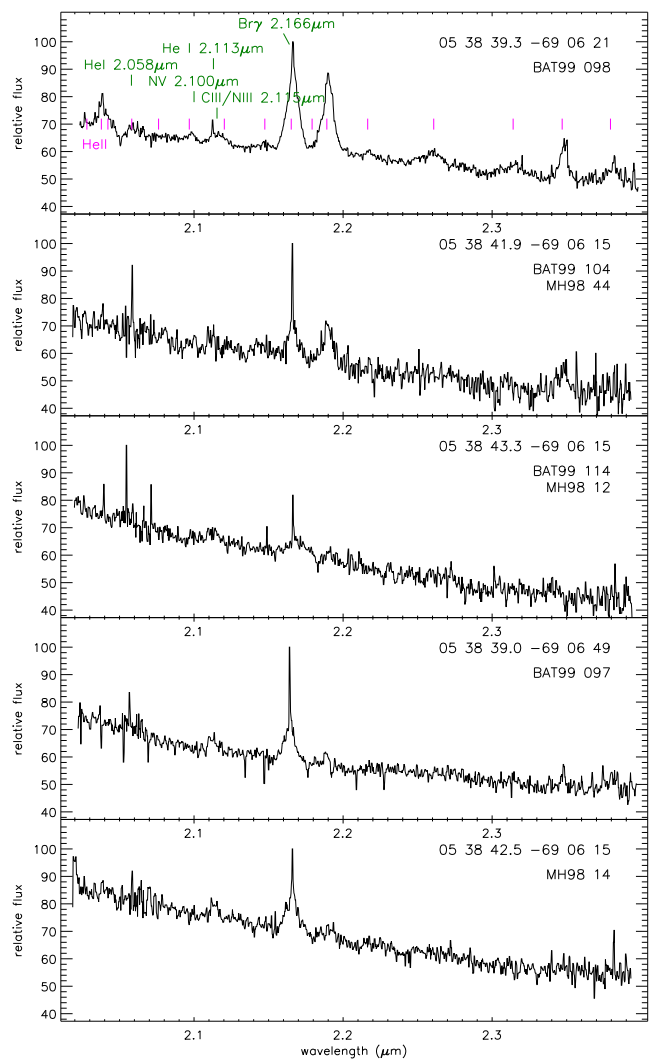}}
\end{center}
\caption{ \label{fig:r136wr}
Spectra extracted from the KMOS mosaic, which had 30\,sec exposure time per pointing.
These are of 4 known Wolf-Rayet stars and 1 possible new one. In addition to broad and narrow Br$\gamma$, they show a variety of broad He\,II lines (identified in magenta) typical of WN stars.
Lines and wavelengths are from \cite{fig97}.
}
\end{figure} 

\begin{figure}
\begin{center}
\resizebox{\hsize}{!}
{\includegraphics{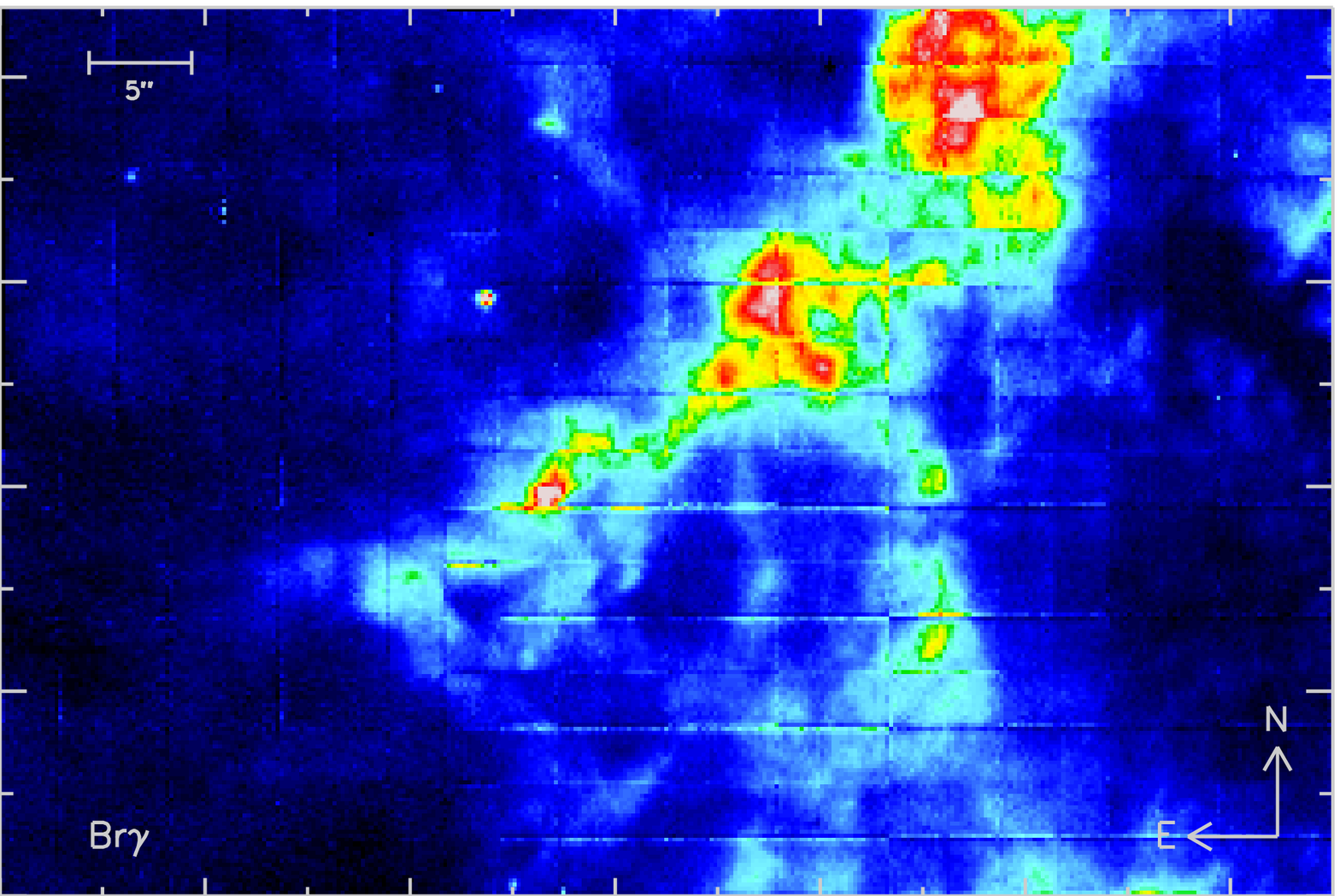}}
\resizebox{\hsize}{!}
{\includegraphics{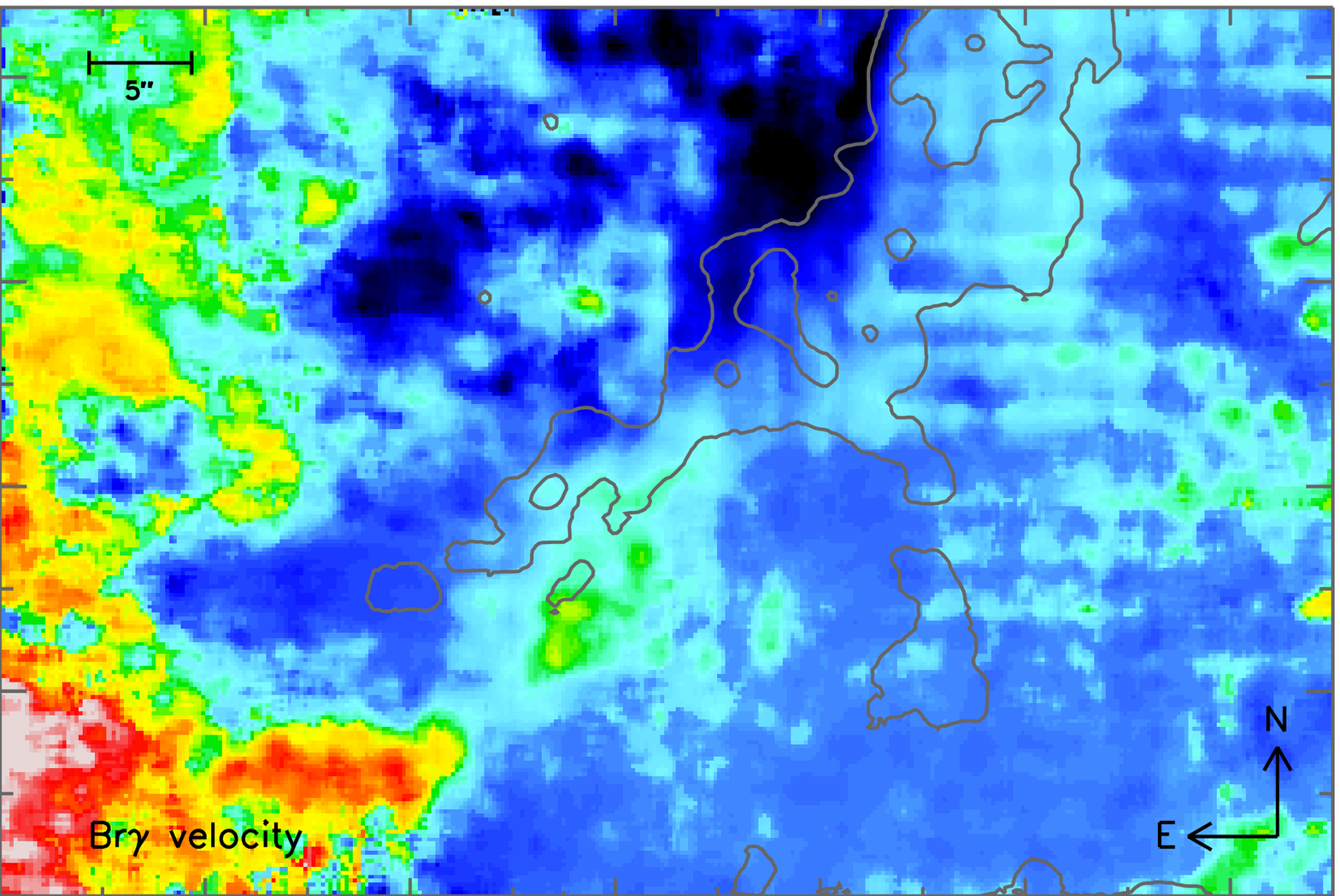}}
\resizebox{\hsize}{!}
{\includegraphics{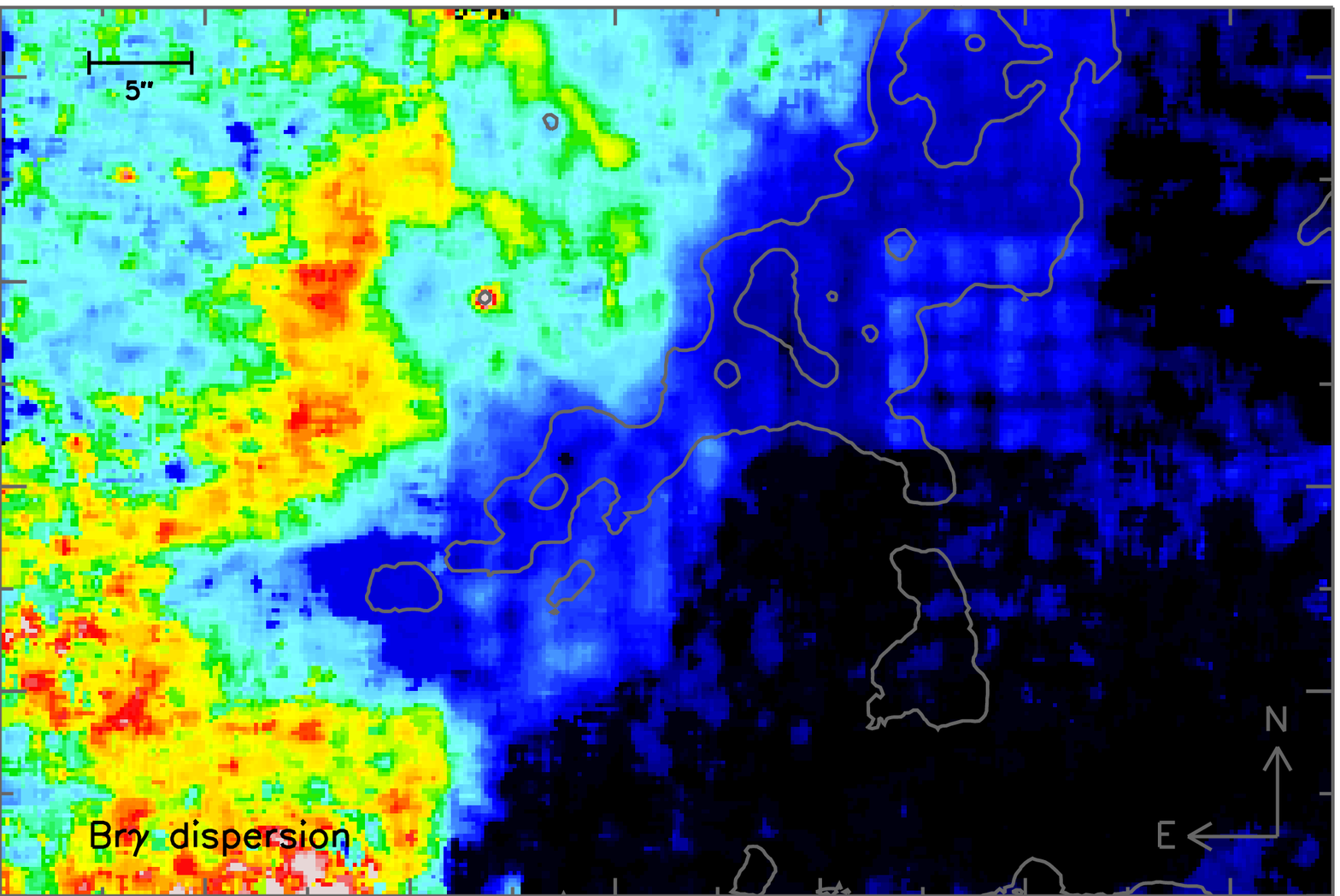}}
\resizebox{\hsize}{!}
{\includegraphics{scalebar.eps}}
\end{center}
\caption{ \label{fig:r136brg} 
Flux (top), velocity (middle) and dispersion (bottom) of the Br$\gamma$ line extracted from the KMOS mosaic.
The velocity is shown in the range 230--310\,km\,s$^{-1}$, and the dispersion in the range 0--70\,km\,s$^{-1}$.
The 2 contours outline the location of the most prominent Br$\gamma$ line emission for reference.
All panels have been drawn on a linear scale using the colour bar shown underneath, with lowest values in black/blue and the highest values in red/white.
}
\end{figure}

\begin{figure}
\begin{center}
\resizebox{\hsize}{!}
{\includegraphics{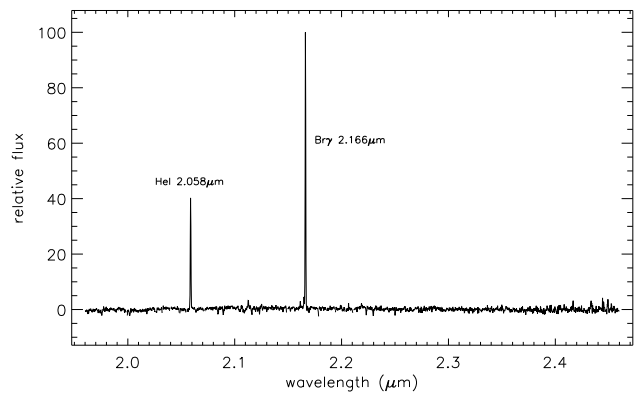}}
\end{center}
\caption{ \label{fig:r136spec} 
Spectrum extracted from the KMOS mosaic of R\,136. 
This is from a region of bright narrow Br$\gamma$ emission at 2.17\,$\mu$m, to the north of the field. It shows also the He\,I line at 2.06\,$\mu$m, but no other features.
}
\end{figure} 

R\,136 is a well-known super star cluster in the 30\,Doradus (Tarantula Nebula) region of the Large Magellanic Cloud.
It is the focus of numerous observational and theoretical studies because it is the most massive young resolved cluster. 
As such it is yielding insights into the formation, environment, content, and evolution of star clusters as well as of starbursts.
It provides an excellent example of what KMOS can do, and also of some of the issues of which one needs to be aware in the processed data (at the same time this shows that further improvements to the pipeline would still be beneficial).
The data have been processed using only the kmo\_sci\_red recipe, and then the extracted maps have been cleaned to remove deviant pixels.
This field was observed in the K-band in mapping mode using 30\,sec exposures, to provide a large mosaic made of 384 tiles.

Fig.~\ref{fig:r136im} shows three maps extracted from the mosaic cube, of a region approximately $60\arcsec\times40\arcsec$ to the south of the cluster core (which is about 10\arcsec\ beyond the top of the field towards the left side).
The top map is the 2.2--2.3\,$\mu$m continuum, which is a wavelength range free of the strongest line emission.
This reveals the stars, the number density of which increases to the north-east.
It also shows variations of the background level within and between the IFU fields, which can make reliable photometry of faint continuum sources difficult to achieve.
A more sophisticated background adjustment algorithm than that applied here could improve this situation.
The map of broad He\,II shows five stars with detections of this line at 2.19\,$\mu$m, the spectra of which are drawn in Fig.~\ref{fig:r136wr}.
A spectrum of the brightest of these stars, closest to the centre of the field in the He\,II image, is drawn in the top panel of the figure.
It indicates that there are numerous He\,II lines with FWHMs of about 1000\,km\,s$^{-1}$, typical of Wolf-Rayet stars as is clear, for example, from K-band spectra of WN stars in the Galactic Center \citep{mar07}, the Arches cluster \citep{mar08}, and also R\,136 itself \citep{cro10}.
These are also prominent in a map of broad Br$\gamma$ flux, which hints that there may be another 2 stars with some broad Br$\gamma$ emission, just an arcsec east of two of the magenta arrows drawn on the mosaics.
However, in the short 30\,sec integration time of these data, the signal-to-noise is too low to confirm these.

All five stars are listed in either the catalogue of Wolf-Rayet stars in the Large Magellanic Cloud by \cite{bre99} or in \cite{mas98} (for which we designate names as MH98 followed by their numbering).
The identifications in both catalogues, together with coordinates, have been given with the spectra Fig.~\ref{fig:r136wr}.
They are all classified as O4\,If+, O3\,If*, WN6-A, or WN7-A stars.
These are consistent with the broad Br$\gamma$ emission, and in some cases broad He\,II lines.
The broad Br$\gamma$ map shows 2 additional tentative detections of WN stars, but their spectra are too noisy, and the equivalent width of the line to low, to attempt any classification.
One of these stars can be identified with MH98~11, which \cite{mas98} class as O4\,If+. 
The other is $\sim$1\arcsec\ east of MH98~14, and does not appear in either of the catalogues mentioned above.
Indeed, it has no obvious counter-part in the {\em WFPC2} image from {\em HST}; and so because the star lies exactly in the join of the tiles in the KMOS mosaic, we conclude that the broad Br$\gamma$ detection in this case may be spurious.

The top map in Fig.~\ref{fig:r136brg} is of the narrow Br$\gamma$ line emission at 2.17\,$\mu$m, created using the LINEFIT code described in Appendix~B of \cite{dav11}.
A spectrum extracted from a bright part of the line emitting region near the north edge of the image is presented in the top panel of Fig.~\ref{fig:r136spec}, and shows also 
narrow He\,I 2.06\,$\mu$m emission but no other features.
While there is a remarkable amount of detail in the Br$\gamma$ image, a number of limitations are immediately clear.
The edge effects between the separate pointings can be severe, and make it difficult to see, for example, that the bright emission at the north edge of the mosaic is actually a ring.

The map was generated by finding, at each spatial location, the Gaussian that, when convolved with the instrumental line profile, provides the best fit to the observed line.
As such, it also yields velocity and dispersion maps.
These are shown in the lower panels of Fig.~\ref{fig:r136brg} and warrant some discussion.
The mosaic was reconstructed using the spatial and spectral flexure corrections described in Section~\ref{sec:compensate}.
Without these, the velocity map would have severe discontinuities; 
applying the corrections (as is the default option) is mandatory.
However, this still leaves variations of up to $\sim$10\,km\,s$^{-1}$ between and within some specific IFUs.
A similar effect is apparent in the dispersion map, where the impact is more severe because of the spatial and spectral variations in instrumental line profile.
The map presented here was extracted using a single line profile in LINEFIT, and has been approximately corrected for the most severe variations (around 10--15\,km\,s$^{-1}$) between the three instrument segments.
It is clear that a valid dispersion can only be recovered using the appropriate spectral profile at each location, rather than assigning a single spectral profile to the whole dataset.
Nevertheless, the maps show that the short section of filament of ionised gas extending across the field from northwest to southeast is dynamically cool, and that the kinematics of the gas towards the cluster centre (to the northeast of the field) are much more complex.

%
\section{Conclusions}
\label{sec:conc}

KMOS is a fully cryogenic multi-IFU spectrograph that has been commissioned at the VLT, and has begun observing operations during 2013.
The instrument produces data that are complex in their raw format.
We have discussed how these data are processed by the pipeline from an algorithmic perspective, and presented one set of results as an example.

The pipeline has been successfully tested on a variety of different data sets taken over several observing runs, but is also still being developed further.
At the current time, the kmo\_multi\_reconstruct recipe is under revision so that it can take into account the flexure compensation described in Section~\ref{sec:compensate} that is already available in kmo\_sci\_red, and to make it a full science pipeline in its own right.

Another part of the continuing development process is to implement alternative interpolation schemes.
The drizzle method \citep{fru02} has been applied very successfully to 2-dimensional data, especially data that are undersampled.
Related to this is pixel interlacing, a special case of drizzle, which avoids correlating the noise in neighbouring re-sampled pixels.
It is planned that a method based on drizzle should be added to the pipeline in the near future, specifically with the multi-reconstruct concept (Sec.~\ref{sec:multirecon}) in mind.

A third aspect to the development is related to the pairing of object and sky frames/IFUs.
The pipeline does this automatically and there is no easy way for the user to influence the pair allocations.
At the current time, an option to write out a `pairing allocation table' is being included.
In the future, it is intended that a user should able to feed an edited version of the table back into the pipeline, giving the user complete freedom over this part of the data handling.
This capability would provide a simple way to change which sky frame is assigned to any object frame, to allow a sky frame from a different IFU (of either the same or a different exposure) to be used, and even to over-ride the object/sky classifications in the header.

More information about the instrument, and the software needed to use the pipeline, can be accessed through the ESO webpages.
The instrument page at ESO is at\vspace{1mm}\\
\begin{small}
http://www.eso.org/sci/facilities/paranal/instruments/kmos/\vspace{1mm}\\
\end{small}
The software can also be accessed via the wiki maintained by MPE at\vspace{1mm}\\
\begin{small}
https://wiki.mpe.mpg.de/KMOS-spark/\vspace{1mm}
\end{small}

There are several viewers that are compatible with KMOS data.
One that specifically can display datacubes for the 24 IFUs simultaneously, is QFitsView, accessible from\vspace{1mm}\\
\begin{small}
http://www.mpe.mpg.de/~ott/QFitsView/\vspace{1mm}\\
\end{small}
and another is the ESO 3D visualisation tool (CASA viewer) that was developed for the Atacama Large Millimeter/submillimeter Array (ALMA) but is able to display datacubes from other integral field spectroscopy instruments including KMOS\vspace{1mm}\\
\begin{small}
http://www.eso.org/sci/software/pipelines/\vspace{1mm}\\
\end{small}

\begin{acknowledgements}
We thank the staff of ESO in Garching and at the VLT for their dedicated support during the commissioning of the instrument, and for their thorough testing of the pipeline during its early phases.
The results presented are based on public data released from the KMOS commissioning observations at the VLT Antu (UT1) telescope.
We are also very grateful to both Fabrice Martins and Chris Evans for their helpful advice about the stellar spectra and WN identifications.
And we thank the referee for making a number of useful suggestions.
This paper is dedicated to the memory of Carlo Izzo, who was always enthusiastic about the pipeline but never saw its completion.
\end{acknowledgements}


\end{document}